\documentclass[10pt]{article}

\usepackage{moreverb}

\usepackage{geometry}
\usepackage[colorlinks,bookmarksopen,bookmarksnumbered,citecolor=red,urlcolor=red]{hyperref}
\usepackage{breakurl}

\usepackage[T1]{fontenc}        

\usepackage[utf8x]{inputenc}    

\usepackage{pifont}

\usepackage{float}
\usepackage{subfig}
\usepackage{caption}
\usepackage[export]{adjustbox}
\usepackage{afterpage}

\usepackage{mathptmx}
\usepackage{soul}\setuldepth{article}
\usepackage{ucs}                
\usepackage{amsmath}            
\usepackage{amsfonts}
\usepackage{amssymb}
\usepackage{amsthm} 
\usepackage{mathtools} 
\usepackage{thmtools}
\usepackage{dsfont}
\usepackage{color}
\usepackage{graphicx} 
\usepackage{textcomp}            
\usepackage{palatino}            
\usepackage{eulervm}            

\usepackage{varioref}            
\usepackage{cleveref}             
\usepackage{fancyhdr}
\usepackage{multirow}
\usepackage{rotating}
\usepackage{xr} 
\usepackage{xcite}
\usepackage{setspace} 
\usepackage{tabularx}

\newcommand{\R}{{\mathbb{R}}}

\newcommand{\N}{{\mathbb{N}}}        

\usepackage{titlesec}
\titleformat{\subsection}
  {\normalfont\large\bfseries\itshape}  
  {\thesubsection}                   
  {0.6em}                            
  {}                                 

\begin{document}
\pagenumbering{gobble}
\def\spacingset#1{\renewcommand{\baselinestretch}%
	{#1}\small\normalsize} \spacingset{1}

\begin{center}

\begin{Large}
\textbf{How to Simulate Realistic Survival Data? A Simulation Study to Compare Realistic Simulation Models}
\end{Large}

\vspace{0.7cm}

Maria Thurow${}^{1,2}$, Ina Dormuth${}^{1}$, Christina Sauer${}^{3}$,\\ 
Marc Ditzhaus${}^{4}$ and Markus Pauly${}^{1,2}$\\[2ex]
\noindent${}^{1}$ Department of Statistics, TU Dortmund University, Dortmund, Germany\\
\noindent${}^{2}$ Research Center Trustworthy Data Science and Security, UA Ruhr, Dortmund, Germany\\
\noindent${}^{3}$ Institute for Medical Information Processing, Biometry, and Epidemiology, Faculty of Medicine, Ludwig-Maximilians-University, Munich, Germany\\
\noindent${}^{4}$ Faculty of Mathematics, Otto-von-Guericke-University, Magdeburg, Germany

\vspace{0.5cm}

\begin{abstract}
In statistics, it is important to have realistic data sets available for a particular context to allow an appropriate and objective method comparison. 
For many use cases, benchmark data sets for method comparison are already available online. However, in most medical applications and especially for clinical trials in oncology, there is a lack of adequate benchmark data sets, as patient data can be sensitive and therefore cannot be published.
A potential solution for this is simulation studies.
However, it is sometimes not clear, which simulation models are suitable for generating realistic data. 
A challenge is that potentially unrealistic assumptions have to be made about the distributions. Our approach is to use reconstructed benchmark data sets 
as a basis for the simulations, which has the following advantages: the actual properties are known and more realistic data can be simulated. There are several possibilities to simulate realistic data from benchmark data sets. We investigate simulation models based on kernel density estimation, fitted distributions, case resampling and conditional bootstrapping. 
In order to make recommendations on which models are best suited for a specific survival setting, we conducted a comparative simulation study.
Since it is not possible to provide recommendations for all possible survival settings in a single paper, we focus on providing realistic simulation models for two-armed phase III lung cancer studies. To this end, we reconstructed benchmark data sets from recent studies. We used the runtime and different accuracy measures (effect sizes and p-values) as criteria for comparison.\\[2ex]
\end{abstract}

\vspace{0.5cm}
\end{center}

\noindent Corresponding author: Maria Thurow, Department of Statistics, TU Dortmund University, 44221 Dortmund, Germany. E-mail: maria.thurow@tu-dortmund.de.

\newpage
\pagenumbering{arabic}

\def\spacingset#1{\renewcommand{\baselinestretch}%
	{#1}\small\normalsize} \spacingset{2}
\section{Introduction}\label{sec:introd}
Model comparison and model selection are important tasks in statistical analyses.
As described by Friedrich and Friede \cite{Friedrich2023Benchmark}, benchmarking and simulation studies are two different approaches to use for these tasks. The idea of benchmarking is to compare methods or models based on multiple neutrally chosen real world data sets regarding a specific performance criterion.
Benchmarking aims at receiving ``empirical evidence'' from model comparison on several real world data sets. This has the advantage of providing a realistic approximation of the performance of the model in subsequent analyses \cite{Boulesteix2017}. It is important to follow existing 
guidelines for benchmarking (for example, choosing the evaluation criteria and methods in a neutral way) to ensure a high scientific quality and to obtain a realistic and unbiased comparison of the models \cite{Boulesteix2017}.

In contrast, 
simulation studies 
allow generating an infinite amount of data even for very specific and potentially complicated settings. For model comparison, simulation models can be additionally used to compare different models when assumptions are violated or when it is difficult to find an adequate data set.
Another advantage of simulation models is that the distribution and other aspects of the simulated data are well-known which allows a comparison of the performance of models for slightly different (theoretical) settings \cite{Morris2019}. A disadvantage of ``simply'' simulating the data is that there might be 
unrealistic assumptions regarding the data sets. This could result in a very good performance of a certain model for a specific simulated data set. However, when considering another (for example neutrally chosen benchmark) data set, the model might perform worse. This could lead to scientists, intentionally or unintentionally, choosing simulation models or assumptions supporting a new model they want to present or a model they prefer. The results of the comparison study therefore might be biased as for example described by Nie{\ss}l et al. \cite{Niessl2021}. 

In oncology, model comparison (e.g., to select the proper method for a new analysis) is complicated by the fact that not many data sets are publically available since for example patient data is sensitive personal data and can therefore not be published easily. In fact, Hamilton et al. \cite{Hamilton2022DataAvailability} found out that from the cancer-related publications on the online platform PubMed \cite{PubMed} only 59 out of 309 studies published in 2019 declared their data to be publicly available. From these 59 publications, only one single data set fulfilled all of the FAIR Data Principles, which in short means that the data should be ``Findable, Accessible, Interoperable, and Reusable'' \cite{Wilkinson2016FAIR}.  Thus there is a lack of suitable real world benchmark data sets. As a solution,  oftentimes simulated data is used for model comparison. This raises the question of how to simulate realistic survival data for a neutral model comparison?
Simulating realistic survival data, such as from parametric distributions, has been explored by various researchers before \cite{Brilleman2021SimSurv, crowther2013simulating}.

The present paper aims at answering this question focusing on two-armed phase III lung cancer studies. We thereby combine ideas from benchmarking with simulation studies. To this end, we first need an objective selection of realistic and relevant benchmark data sets. Due to the availability issue mentioned above, we provide \textit{reconstructed} benchmark data sets from recent studies on lung cancer patients.
To do so, we first performed an extensive literature search for adequate studies. Based on different criteria, clinical studies from oncology are selected to be reconstructed and used as benchmark data sets. For the reconstruction, we applied the widely used algorithm of Guyot et al. \cite{Guyot2012}.
Based on each of these benchmark data sets, we build four different types of simulation models which are either based on kernel density estimation, fitting of parametric distributions or two different bootstrap approaches.
These are compared in a simulation study regarding their ability to simulate realistic survival data. To quantify how realistic the simulated data sets are, we analyze different metrics of the simulated data sets and compare them to the values reported in the respective publications of the studies.

This paper is structured as follows: In Section \ref{sec:data_sets}, the selection procedure of the studies is described and the data sets as well as the reconstruction algorithm are presented. The compared simulation models are introduced in Section \ref{sec:simu_mods}. In Section \ref{sec:simulation}, the simulation setup is described and the results are presented in Section \ref{sec:results}. Lastly, the simulation and its results are summarized in Section \ref{sec:conclusion}.
\section{Data Sets}\label{sec:data_sets}
The studies used for the analysis were selected according to previously defined criteria which are described below and are searched for via the online platform PubMed \cite{PubMed}. We looked for phase III clinical trials from oncology, published before November 2022, in which the Kaplan-Meier (KM) estimator is reported. This information is necessary for the reconstruction of the data (see below). An additional criterion was that the $p$-value of the logrank test had to be reported in the studies since this is used as a benchmark criterion for the simulation. Applying the criteria resulted in 290 potential studies.

From these, studies were selected by considering additional criteria.
Besides the KM estimator, the number at risk should be reported in the publication as well, since it is needed to apply the reconstruction algorithm. Additionally, the total number of events should be reported. Since there might be differences regarding the reconstruction quality for the different cancer types, only studies on lung cancer patients were used as this is a gender-independent and widespread cancer type for which many studies are available.  In 2020, it was the second most common cancer type (following breast cancer) with over two million newly diagnosed cases \cite{Sung2021cancerstatistics}. To increase the accuracy of the data reconstruction, we limited our simulation to two-armed trials (for example, a specific therapy and a placebo group) in which the reported KM curves of the two groups do not cross more than twice. Since the accuracy of the reconstruction depends on the resolution of the plots used, a high resolution is necessary. Considering the additional criteria resulted in ten studies.
Based on the KM curves, these studies can be divided into three groups: non-crossing KM curves (with no visible late effect), non-crossing KM curves with a visible late effect in the data, and crossing KM curves. From the ten studies, six can be assigned to the first group, one to the second, and three to the third. 
In order to have a similar number of studies with similar sample sizes in the three groups, three of the six studies are selected for the group of studies with crossing KM curves.

Non-crossing KM curves with no late effect are characterized by a higher survival probability in one group compared to the other group for the whole study time. The difference between the curves typically does not change much over time.
This is typical for proportional hazard rates.  The studies of Spigel et al. \cite{Spigel.2022}, Wei et al. \cite{Wei.2020} and Cordeiro de Lima et al. \cite{CordeirodeLima.2018} are assigned to this group.

A late effect in the data is present when the gap between the KM curves grows after a certain time point. Additionally, in this case, only studies that have non-crossing KM curves are considered. In Wu et al. \cite{Wu.2015}, the effect of Erlotinib is compared to the combination of Gemcitabine and Cisplatin (GP).

Crossing survival curves indicate a change in the survival probability, in such a way that the group with a previously higher survival probability has a lower survival probability compared to the other group. Three data sets of this group were used for the simulation (Yoshioka et al. \cite{Yoshioka.2019}, Liang et al. \cite{Liang.2017} and Seto et al. \cite{Seto.2020}). 

\begin{figure}[!h]
\centering
\subfloat[Non-crossing KM curves (Cordeiro de Lima et al. \cite{CordeirodeLima.2018})]{\label{fig:recon_noncross}\includegraphics[width = 0.41\linewidth, valign = b]{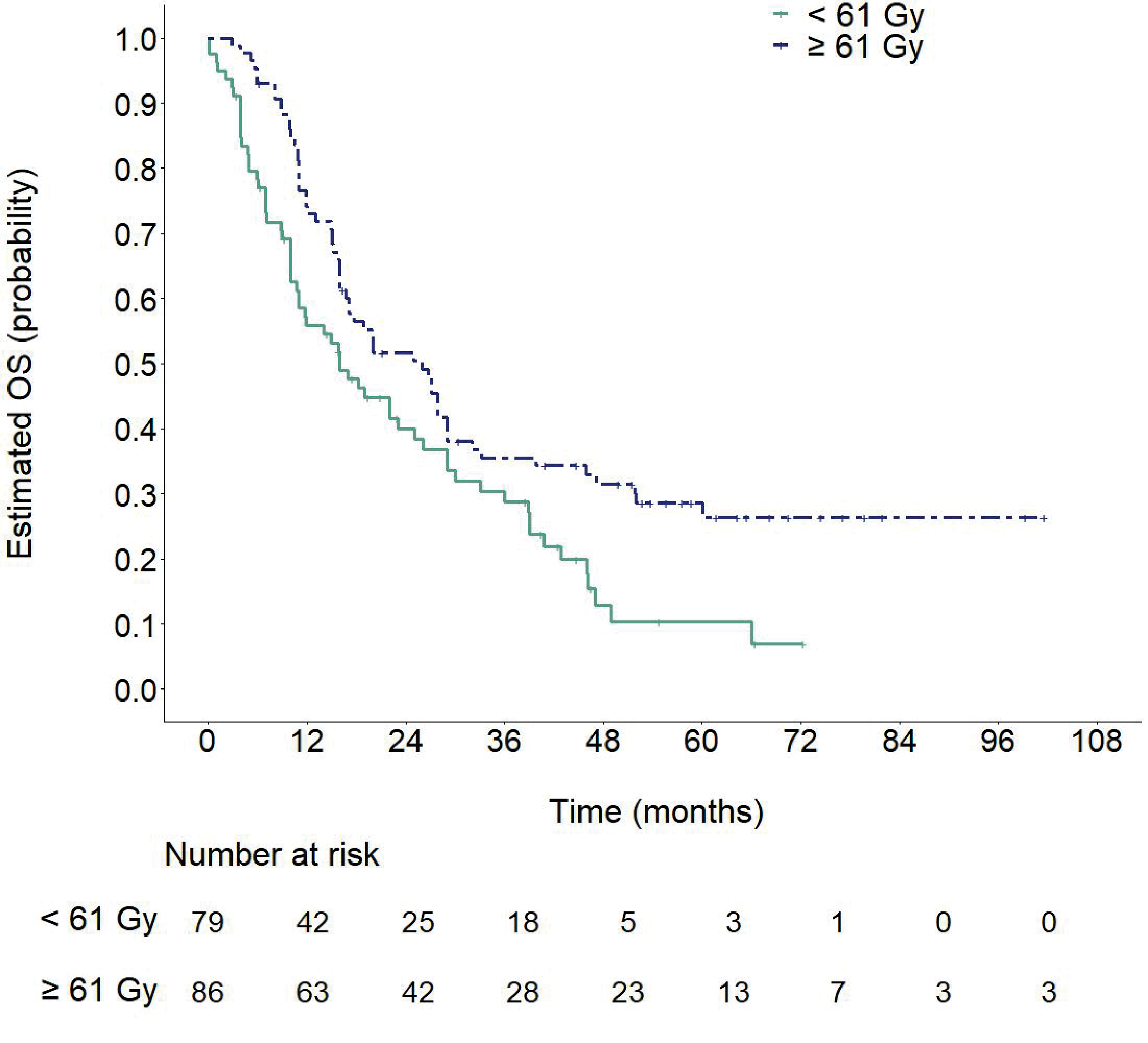}}
\hfill
\subfloat[Non-crossing KM curves with a late effect (Wu et al. \cite{Wu.2015})]{\label{fig:recon_noncrosslate}\includegraphics[width = 0.44\linewidth, valign = b]{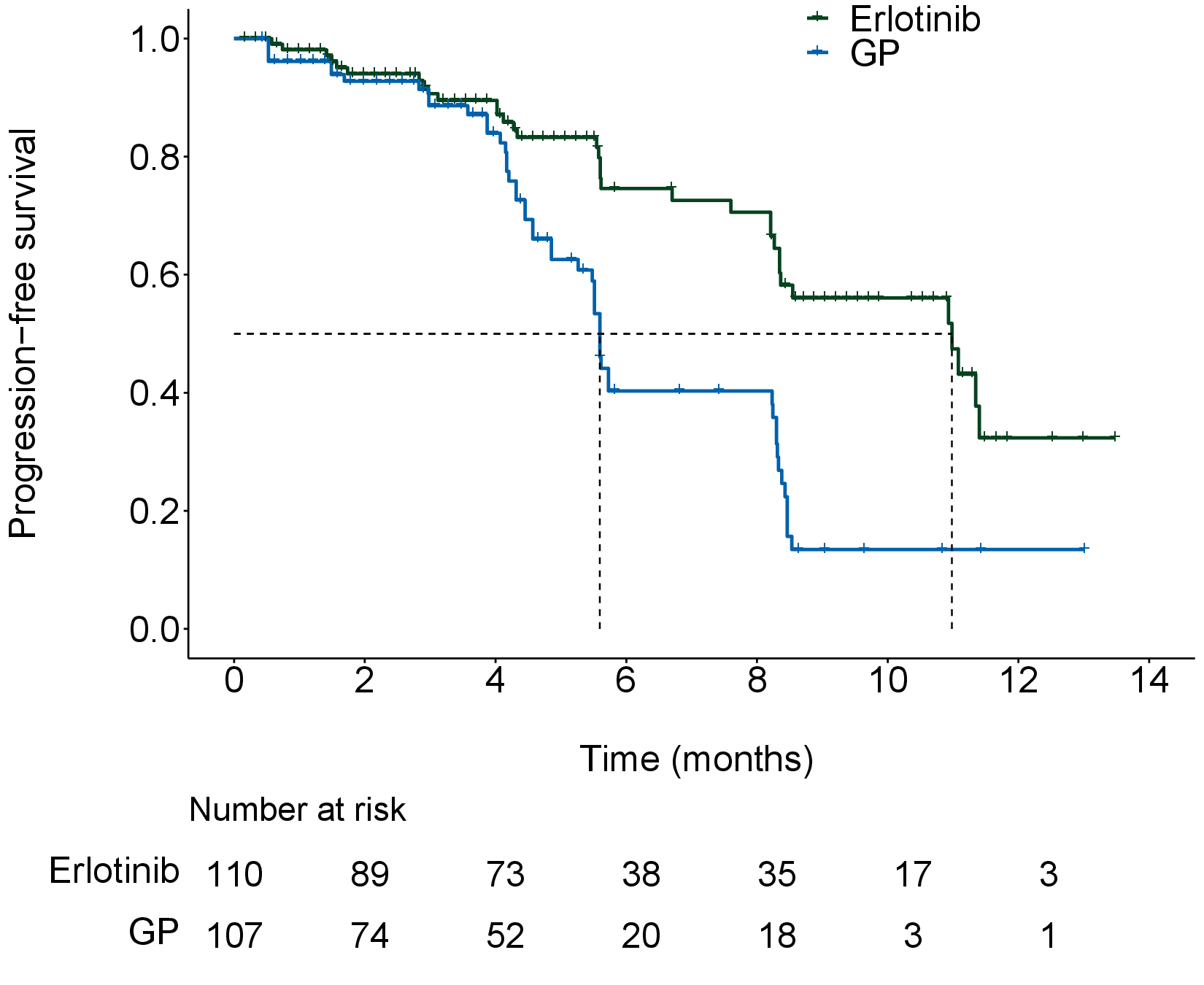}}
\vphantom{\includegraphics[width=0.41\linewidth,valign=b]{Pictures/km_curves/lima_recon.eps}}\\[2ex]
\subfloat[Crossing KM curves (Liang et al. \cite{Liang.2017})]{\label{fig:recon_cross}\includegraphics[width = 0.4\linewidth]{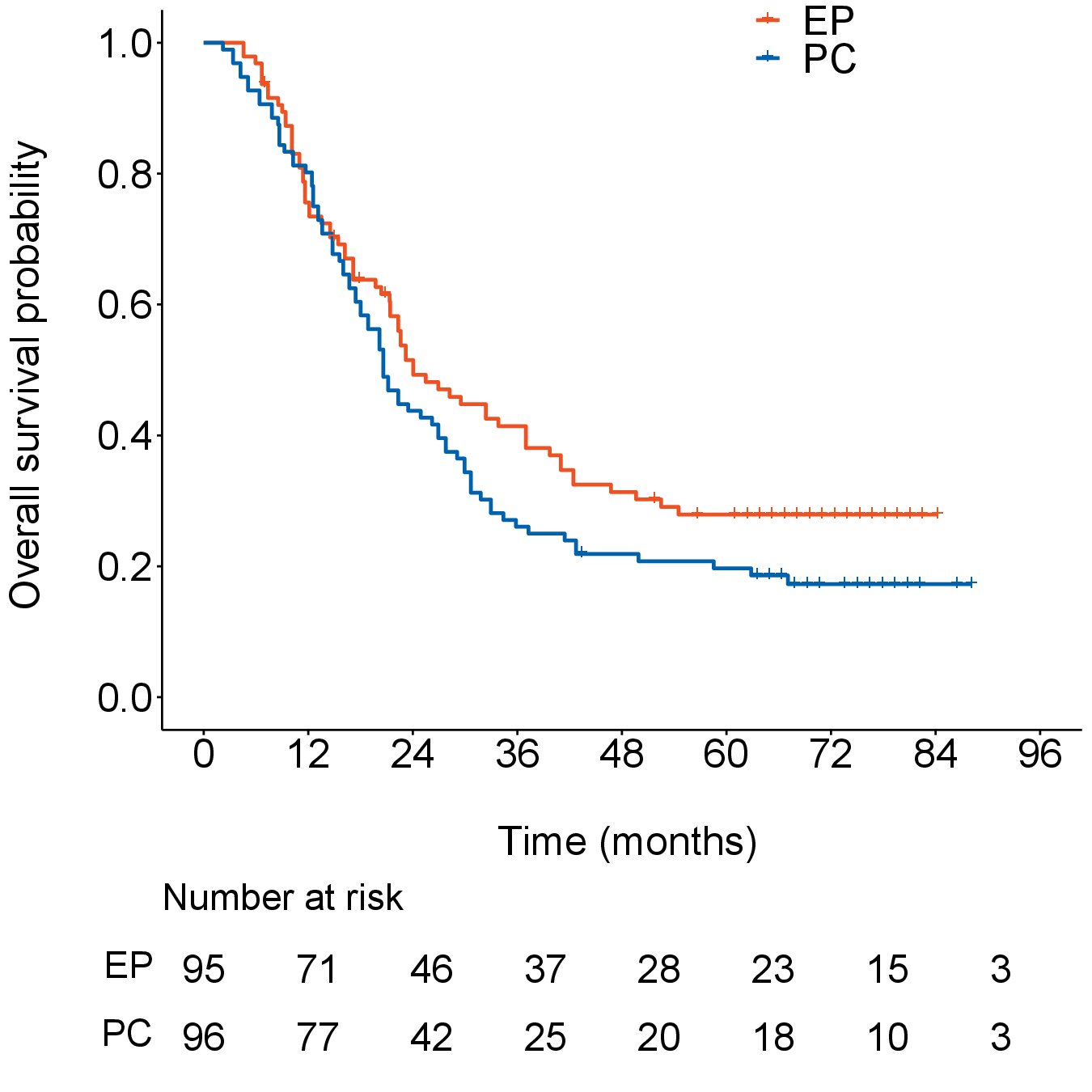}}

\caption{Examples of the KM curves of the reconstructed data sets (one representative per group).}
\label{fig:Reconstruction_representative}
\end{figure}

For each group, Figure \ref{fig:Reconstruction_representative} displays a representative of the KM curves computed from the reconstructed data. The KM curves of the four remaining benchmark data sets can be found in Figure \ref{fig:Reconstruction_appendix} of the Appendix.

The data is reconstructed from the published plots by using the reconstruction algorithm of Guyot et al. \cite{Guyot2012}, which is widely used in this context \cite{Dormuth2022, Spring2020, Berry2017}.
The algorithm uses digitized plots to reconstruct the data. To extract the information from the published plots, the free online tool WebPlotDigitizer \cite{Rohatgi2022} is used. Afterwards, the extracted data has to be formatted in order to be passed to the reconstruction algorithm.
The reconstructed data sets will be made publicly available after approval from the authors of the original studies. 

\section{Simulation Models}\label{sec:simu_mods}
A challenge when using simulation studies is to generate realistic (survival) data. Therefore, we compare four different simulation models to simulate data sets based on the reconstructed data in \texttt{R} \cite{R2022}. In the following, a simulation model based on parametric distributions, a simulation model based on kernel density estimation and simulation models based on two bootstrap methods for simulating censored data are described. 

In this paper, the observations from the considered studies are right-censored and therefore the following methods are only introduced for this type of data. For each observation, there is a theoretically observable event time $T^0>0$ following the distribution with distribution function $F^0$ and a censoring time $C>0$ following the distribution with distribution function $G$. The observed time $T \coloneq \min\{T^0, C\}$ is the minimum of the theoretically observable time and the censoring time of an observation.
The censoring indicator is then defined as 
\[
D \coloneq \begin{cases}
	1, T^0 < C\\
	0, \text{otherwise}.
\end{cases}
\]
The observations are realizations (denoted by lower-case letters) of the random variables and have an index for the respective observation number. The $i^{\text{th}}$ observation ($i=1,\dotsc,n$, $n\in \N$) is a tuple $(t_i = \min\{t^0_i, c_i\}, d_i)$. For $d_i = 0$, the observation is censored, for $d_i=1$, the event is observed. In the following sections, random censoring is assumed and the distributions of the event and censoring times are considered to be independent.

\subsection{Parametric Distributions}
A commonly used method for simulating survival times is based on specific parametric distributions. In survival analysis, for most simulations, the exponential, Weibull, gamma, log-normal, inverse gamma, log-logistic or the Gompertz distribution are used \cite{Bender2005, Dormuth2023ComparativeStudy, lee1997survival}. Additionally the normal, Cauchy and the Gumbel distribution are considered in this paper.

In the following simulation, the parameters of the distributions are estimated by applying maximum likelihood estimation.

To find the most appropriate distribution for a specific data set, the data set itself is used to estimate the parameters for each of the considered distributions. Afterwards, the (one-sample) Cramér-von Mises (CVM) test is applied. The null hypothesis is that the data set follows the respective (fitted) distribution. The distribution with the highest $p$-value is then assigned to the data set since the least amount of indications for distributional deviations can be found for this distribution.
In the simulation, this approach is applied separately to the censoring and event times of each of the two treatment groups.

As for some data, none of the considered distributions seem to fit, an additional mixed distribution is considered. In Figure \ref{fig:hist_mixture} in the Appendix, the histogram of the censoring times of the placebo group from the study of Spigel et al. is shown. Since the distribution seems to be bimodal with some values between 0 and 20 months and a peak at around 60 months, a mixture of a Weibull and a normal distribution seems to fit here. The density function of the mixed distribution $f_{\alpha, \lambda, \mu, \sigma} :\R \mapsto [0,1]$ can be written as
\[
f_{\alpha, \lambda, \mu, \sigma}(x) = 0.2 \cdot f^{\text{weibull}}_{\alpha, \lambda}(x) + 0.8\cdot f^{\text{norm}}_{\mu, \sigma}(x),
\]
where $f^{\text{weibull}}_{\alpha, \lambda}(x)$ is the density function of a Weibull distribution with parameters $\alpha, \lambda >0$ and $f^{\text{norm}}_{\mu, \sigma}(x)$ is the density function of a univariate normal distribution with parameters $\mu \in \R$ and $\sigma >0$.

After finding adequate distributions for the censoring and event time distributions of each group, the data is simulated from the fitted distributions.

\subsection{Kernel Density Estimation}
Another method to simulate data based on a given data set is to use the kernel density estimation (KDE). The density function is estimated by using the function \texttt{kdensity} from the \texttt{R} package of the same name \cite{Moss2020kdensity} with its default settings, i.e. a Gaussian kernel.

Based on this estimated density function, random values were simulated according to 
the accept-reject method as described by Robert and Casella \cite{Robert.1999}.
For the accept-reject method, the random variable $X$ is assumed to follow a distribution with density function $f: \R \mapsto [0,1]$. Additionally, a random variable $U$, uniformly distributed on the interval $[0,f(x)],\ x\in \mathcal{D} \subseteq \R$, is considered, where $\mathcal{D}$ is the range of the random variable $X$. According to Robert and Casella \cite{Robert.1999}, the joint distribution
\[
(X,U) \sim \mathcal{U}\{(x,u): 0 < u < f(x)\}
\]
can be used for the simulation. The density function $f$ can then be written as the marginal distribution of this joint distribution
\[
f(x) = \int_{0}^{f(x)} du.
\]
The fundamental theorem of simulation then ensures that simulating realizations of the random variable $X$ is equivalent to simulating realizations from the random variable $(X,U)$ \cite{Robert.1999}.

In the accept-reject method, two random values are drawn from uniform distributions 
\[
(X,U) \sim \mathcal{U}(\mathcal{D} \times [0,1]).
\]
If the $i^{\text{th}}$ sampled realization, $i=1,\dotsc,n$, $n\in \N$, $(x_i,u_i)$ of $(X,U)$ fulfills the condition $u_i < f(x_i)$, the value $x_i$ can be used as a realization of the desired distribution. Otherwise, new realizations $x_i$ and $u_i$ are drawn. The realizations $x_1,\dotsc,x_n$ then follow the distribution with density function $f$.
Since the distribution functions of the censoring and event times in the treatment groups differ, the data for the censoring and event times of each of the treatment groups is simulated separately.

\subsection{Resampling Methods}
In addition to the two simulation models from above, we also consider two resampling methods for censored data to simulate data sets: Case resampling and conditional bootstrapping \cite{davisonhinkley1997bootstrap}. Both resampling approaches are applied separately to the two treatment arms and described below.

\subsubsection*{Case Resampling}
To simulate a data set of size $n^{\ast}\in \N$ according to the case resampling procedure, $n^{\ast}$ observations $(t_i, d_i)^{\ast}$ ($i=1,\dotsc,n^{\ast}$) are drawn with replacement from the observed tuples $\{(t_1,d_1),\dotsc,(t_n,d_n)\}$ \cite{davisonhinkley1997bootstrap}.
\subsubsection*{Conditional Bootstrapping}
The conditional bootstrap uses additional information to sample the data. For censored observations in the original data, the censoring time is transferred to the simulated data ($c^{\ast}_i = t_i,\ i \in \{j\in \{1,\dotsc,n\}: d_j = 0\}$). For the other observations, the censoring times $c^{\ast}_i,\ i \in \{j\in \{1,\dotsc,n\}: d_j = 1\},$ are sampled as realizations of a random variable with distribution function
\[
\hat{G}_i(c) \coloneq \frac{\hat{G}(c) - \hat{G}(t_i)}{1 - \hat{G}(t_i)} = \hat{P}(C \leq c \mid C > t_i).
\]
This equals the estimated distribution function of the censoring times conditional on $C>t_i$. After sampling or setting the censoring times, the event times $t^{0 \ast}_1,\dotsc,t^{0 \ast}_n$ are sampled with replacement from the theoretically observable event times of the original data $\{t_i : i\in \{j \in \{1,\dotsc,n\} : d_j = 1\}\}.$ To ensure that all observations can be part of a bootstrap sample, the largest values are treated differently. In case the largest observation is censored, the respective event time is set to a time point shortly after the censoring time by adding a small positive value drawn from a uniform distribution ($\mathcal{U}(10^{-30}, 10^{-20})$). For an uncensored observation, the same principle is applied for the censoring time. Otherwise, the largest observation could never be part of the simulated data set.
After that, the observations are set to the minimum of the respective theoretically observable event and censoring times $t^{\ast}_j = \min\left(t^{0\ast}_j, c^{\ast}_j\right)$ and the censoring indicators $d_j\in \{0,1\}$ are set accordingly, $j=1,\dotsc,n$ \cite{davisonhinkley1997bootstrap}.
\section{Simulation Setup}\label{sec:simulation}
The data sets used for the simulation are reconstructed by a widely used reconstruction algorithm \cite{Guyot2012} from the seven clinical studies described in Section \ref{sec:data_sets}. Within the simulation, we compare the four different simulation models described in Section \ref{sec:simu_mods}. It is of interest how well the different models perform in reconstructing realistic survival data. Since there are only few benchmark data sets for the context of survival analysis and especially for clinical trials in oncology, the first step before the simulation is the reconstruction of the data sets. This way we provide the community with benchmark data for lung cancer trials as a nice side benefit. To compare the reconstructed with the original data sets, we calculate the $p$-values of the logrank test as well as the median survival times in the groups for both, the reconstructed and the original, data sets. Additionally, we compare the KM curves of the reconstructed data sets to the ones of the original data.

After reconstructing the data, for each data set and each treatment group of the data sets, a realistic parametric distribution is selected for the first simulation approach. As described in Section \ref{sec:simu_mods}, the suggested distribution for a specific event or censoring time is the distribution leading to the highest $p$-value of the CVM test when estimating the distribution parameters with the maximum likelihood estimation. The suggested distributions for the event and censoring times of the different treatment groups of the seven data sets are shown in Table \ref{tab:distributions}. 
\begin{table}[t]
	\centering
	\caption{Suggested parametric distributions for the different groups of the studies based on the $p$-value of the CVM test.}
	\label{tab:distributions}
	\begin{tabular}{|l|c|c|c|}
		\hline &&&\\[-2.5ex]
		 \multirow{2}{*}{\textbf{Study}} &  \multirow{2}{*}{\textbf{Treatment Group}} & \textbf{Suggested Distribution} & \textbf{Suggested Distribution} \\
		 &&\textbf{Event Times} & \textbf{Censoring Times}\\ \hline
		\multirow{2}{*}{Liang et al.\cite{Liang.2017}} &EP& inverse gamma & inverse gamma \\
		&PC& inverse gamma & inverse gamma \\\hline
		\multirow{2}{*}{Cordeiro de Lima et al. \cite{CordeirodeLima.2018}} &$< 61$ Gy& exponential & inverse gamma \\
  & $\geq 61$ Gy& inverse gamma & mixed\\\hline
		\multirow{2}{*}{Seto et al. \cite{Seto.2020}} & Pem + Bev & inverse gamma & inverse gamma\\
		&Bev& inverse gamma & exponential\\\hline
		\multirow{2}{*}{Spigel et al. \cite{Spigel.2022}} &Placebo& exponential & mixed \\
		&Durvalumab& inverse gamma & mixed\\\hline
		\multirow{2}{*}{Wei et al. \cite{Wei.2020}} & Chemotherapy & inverse gamma & exponential \\
		&MWA/Chemotherapy& exponential & inverse gamma\\\hline
		\multirow{2}{*}{Wu et al. \cite{Wu.2015}} & GP & inverse gamma & log-normal \\
		& Erlotinib & inverse gamma & exponential \\\hline
		\multirow{2}{*}{Yoshioka et al. \cite{Yoshioka.2019}} & CD & inverse gamma & inverse gamma \\
		& Gefitinib & inverse gamma & inverse gamma \\\hline
	\end{tabular}
\end{table}
For most of the groups, the best distribution for simulating the event times according to the methods used is the inverse gamma distribution. For the three groups, the exponential distribution fits better to the event times. For the censoring times, there are other distributions that seem to fit well. For example, for both groups of the study of Spigel et al. and the ``$\geq 61$ Gy'' group of Cordeiro de Lima et al., the mixed distribution of a Weibull and a normal distribution seems to fit better than any single distribution. For the GP group of Wu et al., the log-normal distribution is the most realistic distribution. For most of the other groups, just as for the event times, the inverse gamma distribution fits best. Only few of the groups fit best with an exponential distribution. In this first simulation model, the event and censoring times are simulated from the respective suggested distributions.

For the simulation based on kernel density estimation, the event and censoring times in both groups are considered separately. This gives us four subsets of the data: For each of the two treatment groups, there is one subset for the event times and one for the censoring times. The kernel density estimation and the simulation are then performed on each of the subsets. A similar approach is used for the resampling methods. The difference is that we only consider the two treatment groups separately, giving us two subsets consisting of the event and censoring times in one arm.

In a single simulation step, for each of the four simulation models, a data set is simulated for each of the seven data sets. The sample size of the simulated data sets is equal to the sample size of the respective reconstructed data set since for the conditional bootstrap, the sample size of the simulated data set cannot be larger than the data set itself. For the simulation based on parametric distributions, the distributions shown in Table \ref{tab:distributions} are used. The simulation consists of $10{,}000$ independent iterations of the simulation steps.

\section{Simulation Results}\label{sec:results}

In this section, the results of the data reconstruction and the simulation study are presented. After comparing the quality of the reconstructed data sets, the four simulation models are compared based on their runtime, the $p$-values of the logrank test and the hazard ratios. To identify the different studies, the last name of the first author of each publication (``Lima'' for Cordeiro de Lima et al.) is used in the figures and tables.
\subsection{Quality of the Reconstructed Data Sets}
Before simulating data sets based on the reconstructed data sets, the quality of the reconstruction was evaluated. To do so, the median survival times and the $p$-values of the logrank test can be compared. An overview of the values in the original and reconstructed data sets can be found in Table \ref{tab:recon_comp} in the Appendix.
For most of the data sets, the $p$-values of the original and the reconstructed data do not differ much. However, the $p$-value of the reconstructed data of Seto et al. ($p_{\text{reconstructed}} = 0.171$) differs by more than 0.1 from the $p$-value reported in the study ($p_{\text{original}} = 0.069$). This might have an impact on the results of the simulation study. The median survival times of the reconstructed data sets are similar to the median survival times of the original data sets. The maximum difference here is $0.7$ months for the EP group from Liang et al.
Overall, the quality of the reconstructed data sets can be considered to be good. Only for one study (Seto et al.), larger deviations from the original data can be observed. These differences could not be reduced even by repeating the reconstruction procedure of the data. This might be because the KM curves are very close to each other at the beginning and the end of the study which makes it harder to digitize and reconstruct the data.
Based on these results, all of the reconstructed data sets can be considered to be appropriate for the use as benchmark data sets and can also be used for the simulation.

\subsection{Runtime Comparison}
When comparing different simulation models, it is useful to consider the runtimes of the different models since this can influence later decisions on the model selection. The simulation was performed on LiDO3, the Linux Cluster at the TU Dortmund. Since the simulation for each study was performed on the same cluster node, the runtimes of the different simulation models can be compared without considering the specific resources of the single nodes. In Table \ref{tab:runtimes}, the summary statistics of the runtimes of the different simulation models are shown.
\begin{table}[b]
	\centering
	\caption{Summary statistics of the runtimes (in seconds) for the different simulation models. \label{tab:runtimes}}
	\begin{tabular}{|l|c|c|c|c|c|c|}
		\hline &&&&&&\\[-2.5ex]
		\textbf{Simulation Model} & \textbf{Minimum}&\textbf{1st Quartile} & \textbf{Median} & \textbf{Mean} & \textbf{3rd Quartile} & \textbf{Maximum}\\ \hline &&&&&&\\[-2ex]
		Kernel Density Estimation &0.163&0.478&0.619&1.198&2.559&3.780\\[1ex]
		Case Resampling &0.003&0.003&0.004&0.005&0.006&0.483\\[1ex]
		Conditional Bootstrap &0.006&0.006&0.006&0.010&0.016&0.438\\[1ex]
		Parametric Distributions &0.004&0.004&0.010&0.020&0.024&0.443\\[0.5ex] \hline
	\end{tabular}
\end{table}
Each time, the runtime of the first iteration is removed to avoid a bias in the results (for example, due to compiling of functions in the first iteration). In the table, the runtimes are summarized without differentiating between the studies. Boxplots of the runtimes for the simulation models for the different studies can be found in Figure \ref{fig:runtimes} in the Appendix.
Since the results do not differ much from the overall results, they are not considered separately here. The highest runtimes by far can be observed for the simulation based on kernel density estimation. Simulating one data set by this method takes an average time of 1.198 seconds. For the other models, an average simulation of a data set takes no more than 0.020 seconds. Even the minimum of the runtimes for the kernel density estimation is higher than the third quartile of the other methods and more than 20 times as high as the minimal runtime of the other methods.  The lowest runtimes can be observed for the case resampling method with an average runtime of 0.005 seconds.
The simulation based on parametric distributions has lower runtimes than the conditional bootstrap but is on average half as fast as this method. The median runtime of the simulation based on parametric distributions is just 0.004 seconds higher than for the simulation based on the conditional bootstrap.

When only considering the runtime, the kernel density estimation seems to be the slowest for simulating data sets and should therefore not be used. The runtimes of the other methods are similarly low. Therefore, using one of these methods would be more sensible, in case of similar performances regarding the other criteria.

\subsection{$p$-Values of the Logrank Test}
To test whether or not the survival curves of two (treatment) groups are statistically equivalent, the logrank test can be used \cite{kleinbaum2005introduction}. Thus, the null hypothesis of the logrank test is as follows:
\[
\mathcal{H}_0: S_1(t) = S_2(t)\ \forall\, t>0,
\]
where $S_g:[0,\infty)\mapsto [0,1]$ is the survival function of group $g$, $g\in \{1,2\}$.\\
As described in Section \ref{sec:data_sets}, reporting the $p$-value of the logrank test was a selection criterion for the data sets. 
Therefore, the $p$-value of the logrank test can be used to evaluate the simulation models by comparing the $p$-values of the logrank test of the simulated data with the reported $p$-values from the studies. The differences between the $p$-values of the simulated data sets and the reported values $p_{\text{simulated}} - p_{\text{original}}$ are considered for this comparison. 

\begin{figure}[!ht]
    \centering
    \subfloat[Crossing KM curves]{\label{fig:lr_p_cross}\includegraphics[width=0.48\linewidth]{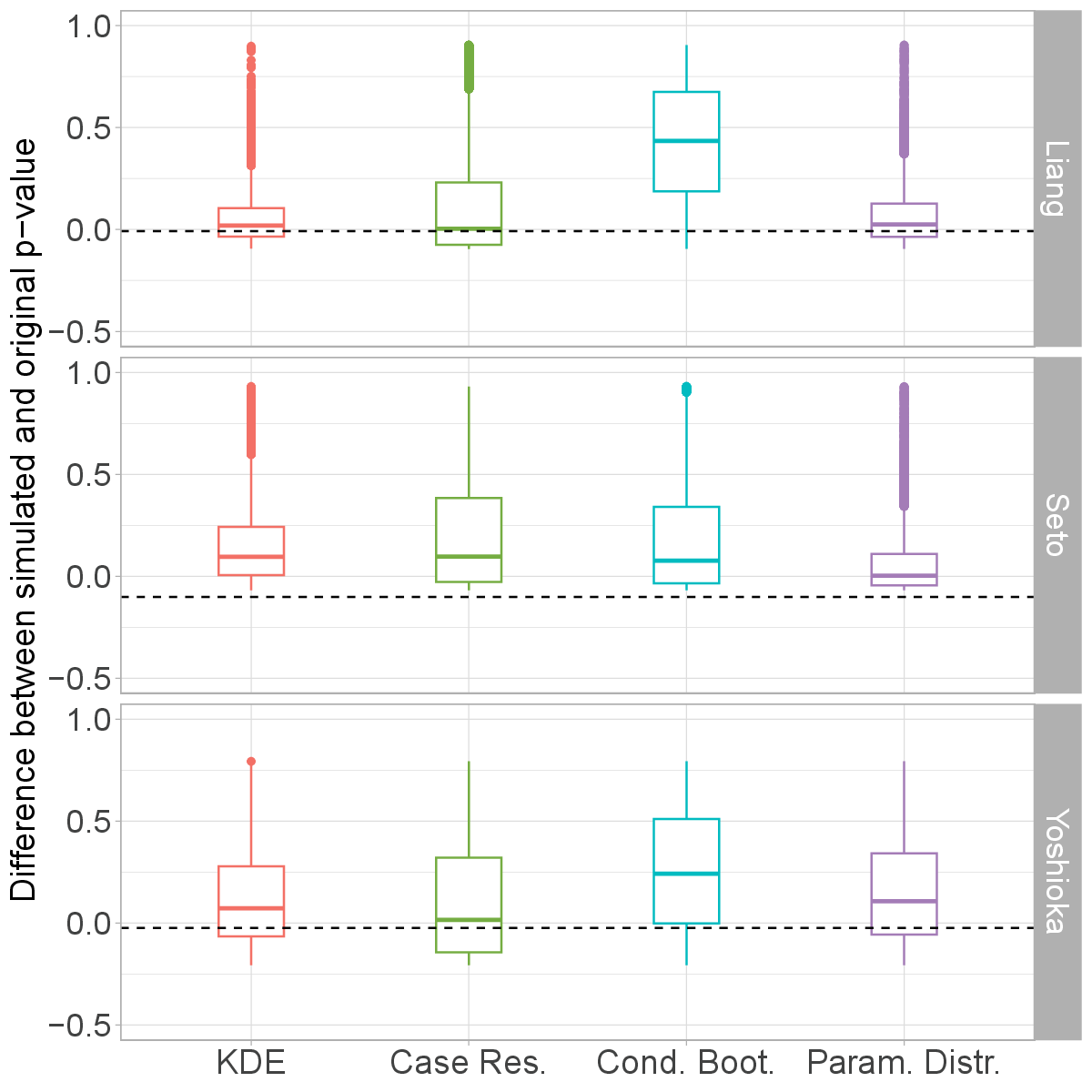}}
    \hfill
    \subfloat[Non-crossing KM curves]{\label{fig:lr_p_non_cross}\includegraphics[width=0.48\linewidth]{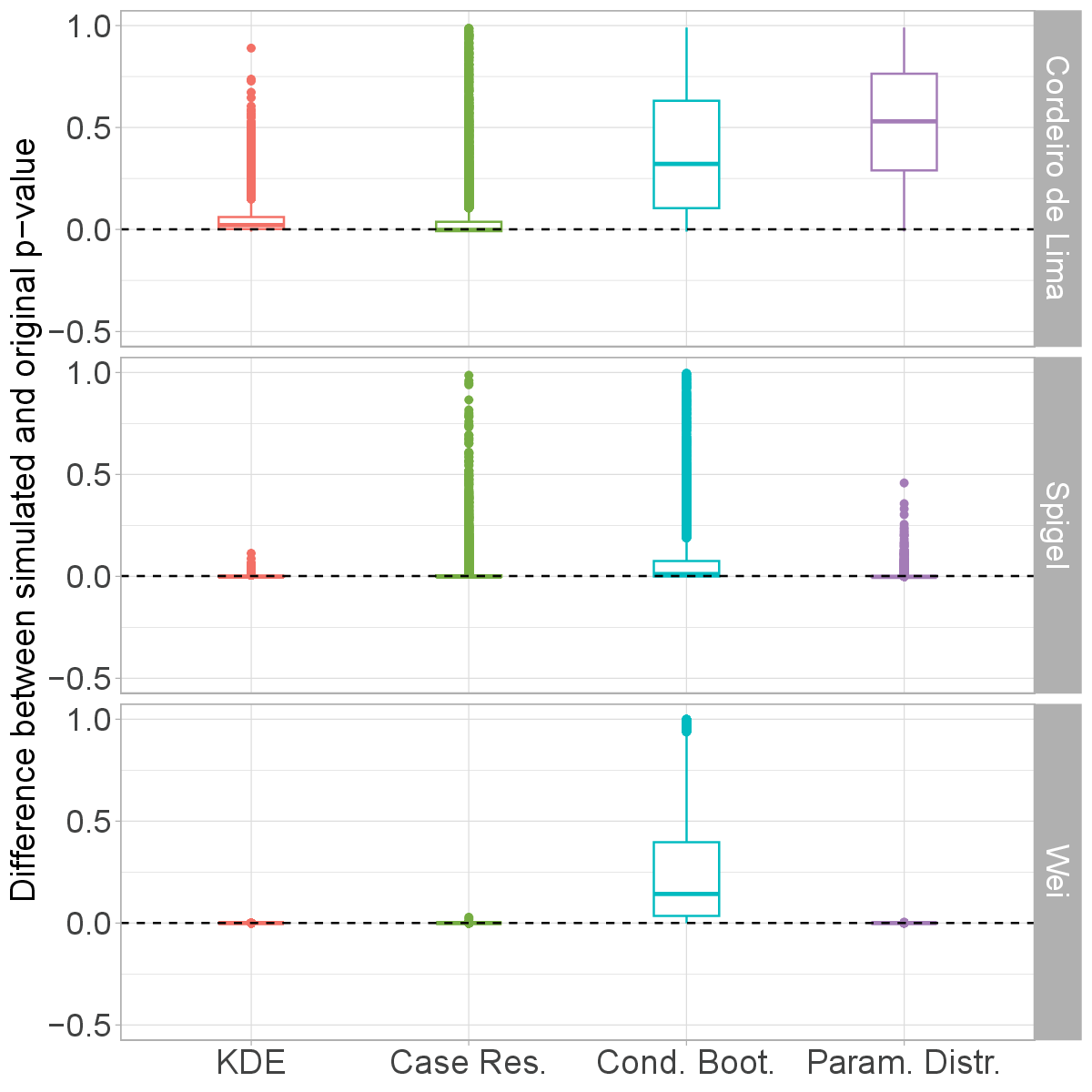}}
    \caption[Boxplots of the differences between the $p$-values of the logrank test of the simulated data sets with crossing and non-crossing Kaplan Meier curves and the $p$-values reported in the studies]{Boxplots of the differences between the $p$-values of the logrank test of the simulated data sets with crossing (left) and non-crossing (right) KM curves and the $p$-values reported in the studies. The dashed line is the difference between the $p$-value of the reconstructed data set and the reported $p$-value.}\label{fig:lr_p_cross_non_cross}
\end{figure}

In Figure \ref{fig:lr_p_cross}, boxplots of the differences between the $p$-values of the logrank test of the simulated data sets and the reported values for the studies, in which the KM curves are crossing once or twice, are displayed. The dashed lines represent the differences between the $p$-values for the reconstructed data sets and the reported $p$-values ($p_{\text{reconstructed}} - p_{\text{original}}$). It can be seen that for two of the three studies, the $p$-value of the reconstructed data set is close to the reported value.
For the third study (Seto et al.), the difference is a bit higher.
This implies that for some studies, the data of the treatment groups is reconstructed well. For the data set from Seto et al., the reconstruction is worse.
For all of the three studies and four simulation models, the median of the $p$-values of the simulated data sets is higher than the $p$-values reported in the study.

The median $p$-value for the data sets simulated by the kernel density estimation for the study by Liang et al. is $0.019$ higher than the reported $p$-value. For the studies by Seto et al. and Yoshioka et al.,  the differences from the reported $p$-values are higher and vary more than for the study by Liang et al. For the data sets simulated by using the case resampling, the $p$-values are close to the reported $p$-values as well while having a higher variability than the values from the data sets simulated by using the kernel density estimation.
The median differences as well as the variability of the $p$-values are very high when using the case resampling.
An explanation for this might be that more similar values from the reconstructed data are included in the bootstrap samples of the groups. When simulating the data sets by using parametric distributions, the median $p$-values are very close to the reported values for all of the three studies in this category. In addition, the variability of the values is very low.

For the studies in which the KM curves are not crossing and there is no late effect, slightly different results can be observed.
The boxplots of the differences between the $p$-values from the logrank test of the simulated data sets and the reported $p$-values are shown in Figure \ref{fig:lr_p_non_cross}.
Similar to the studies with crossing KM curves, the $p$-values of the reconstructed data sets are very close to the $p$-values reported in the studies.

For the simulation based on kernel density estimation, the median $p$-values are very close to the $p$-values reported in the studies and the values do not vary much. The results are the same for the simulation based on the case resampling. When simulating the data sets by using the conditional bootstrap, the results are worse. For the studies by Cordeiro de Lima et al. and Wei et al., the median $p$-value of the data sets simulated by using the conditional bootstrap is higher than the reported $p$-values and the values show a rather high variability. In contrast, for the study by Spigel et al., the variability of the $p$-values is rather low for this simulation method. For the data sets simulated using parametric distributions, the median $p$-values are very close to the reported values for the studies by Spigel et al. and Wei et al. and the values show a very low variability. For the study by Cordeiro de Lima et al., the median value of the differences between the $p$-values of the simulated data and the reported value is very high 
and the values show a high variability. This may indicate that the distributions used for this data set are not similar enough to the original distribution. Another notable aspect of the plots is most of the differences are positive.

When there is an additional late effect in the data, the results are similar to the results for the study from Cordeiro de Lima et al. Therefore, the results are not reported in detail here. The boxplots of the differences between the logrank test $p$-values for the simulated and original data sets are displayed in Figure \ref{fig:lr_p_late} in the Appendix. 

For the groups with crossing KM curves, the simulation based on kernel density estimation, the case resampling and the parametric distributions perform similarly well, although the values for the case resampling vary more than the values of the other methods. When the KM curves are not crossing (with and without a late effect), the simulation based on parametric distributions performs worse for two studies (Cordeiro de Lima et al. and Wu et al.). The simulation models based on kernel density estimation and case resampling perform best for all of the considered studies. For most of the studies, the $p$-values for the simulation based on the conditional bootstrap differ the most from the reported values.
Therefore, regarding the $p$-values of the logrank test, it does not seem sensible to use the simulation based on the conditional bootstrap and parametric distribution. 

For all of the studies, the $p$-values of the simulated data sets show a tendency to be higher than the $p$-values reported in the studies. This might have different reasons. One reason might be a high amount of tied observations in the simulated data sets. In Figure \ref{fig:tie_ratio} in the Appendix, histograms of the tie ratios for the simulated data sets are displayed. It can be seen that for the simulations based kernel density estimation and parametric distributions, no ties are present. For the two bootstrap approaches, for all simulated data sets, more than 50\,\% of the observations are tied. This could be an explanation for the $p$-values being higher for the conditional bootstrap and the case resampling. But the tendency is visible for the other simulation models as well. To further investigate the tendency of the $p$-values, the distribution of the test statistics of the logrank test (in comparison to the test statistic of the reconstructed data) is considered for the simulated data sets. The histograms of the test statistics are displayed in Figure \ref{fig:teststats} in the Appendix. For most of the simulation models and considered studies the distributions of the test statistics of the reconstructed data fit the value of the test statistic observed for the reconstructed data. Only for some simulation settings, the test statistic of the reconstructed data is much higher than most of the test statistics of the simulated data sets. Therefore, this does not explain the tendency of the $p$-values to be higher than the reported $p$-value. To answer the question, other reasons should be investigated in further analyses.

\subsection{Hazard Ratios}
The hazard ratio in a setting with two groups is the ratio between the risk of having an event for a person being in one group and the risk of the same person having an event when being in the other group. It is often used to interpret the effect of being in a specific group on the survival time \cite{klein2003survival}.
Since the hazard ratio is a commonly used and relevant measure in survival analysis, we use it here as another measure to compare the simulation models. As for the $p$-values, the differences between the hazard ratios for the simulated data sets and the hazard ratios reported in the studies are considered to make the comparison between the methods easier. In all of the figures, the differences between the hazard ratios of the reconstructed data sets and the reported hazard ratios are included as dashed lines.

\begin{figure}[!b]
    \centering
    \subfloat[][Crossing KM curves]{\label{fig:hr_cross}\includegraphics[width=0.48\linewidth]{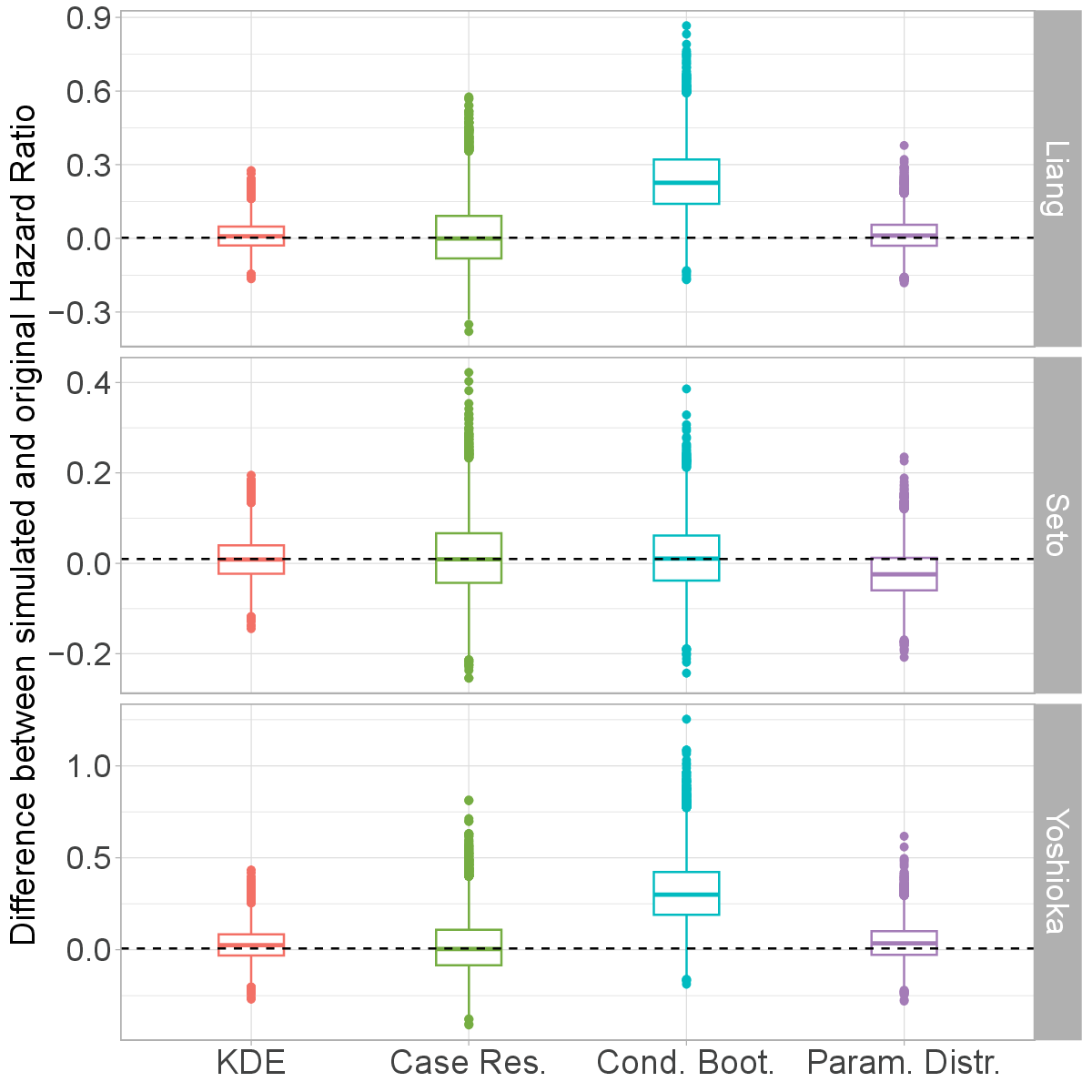}}
    \hfill
    \subfloat[][Non-crossing KM curves]{\label{fig:hr_non_cross}\includegraphics[width=0.48\linewidth]{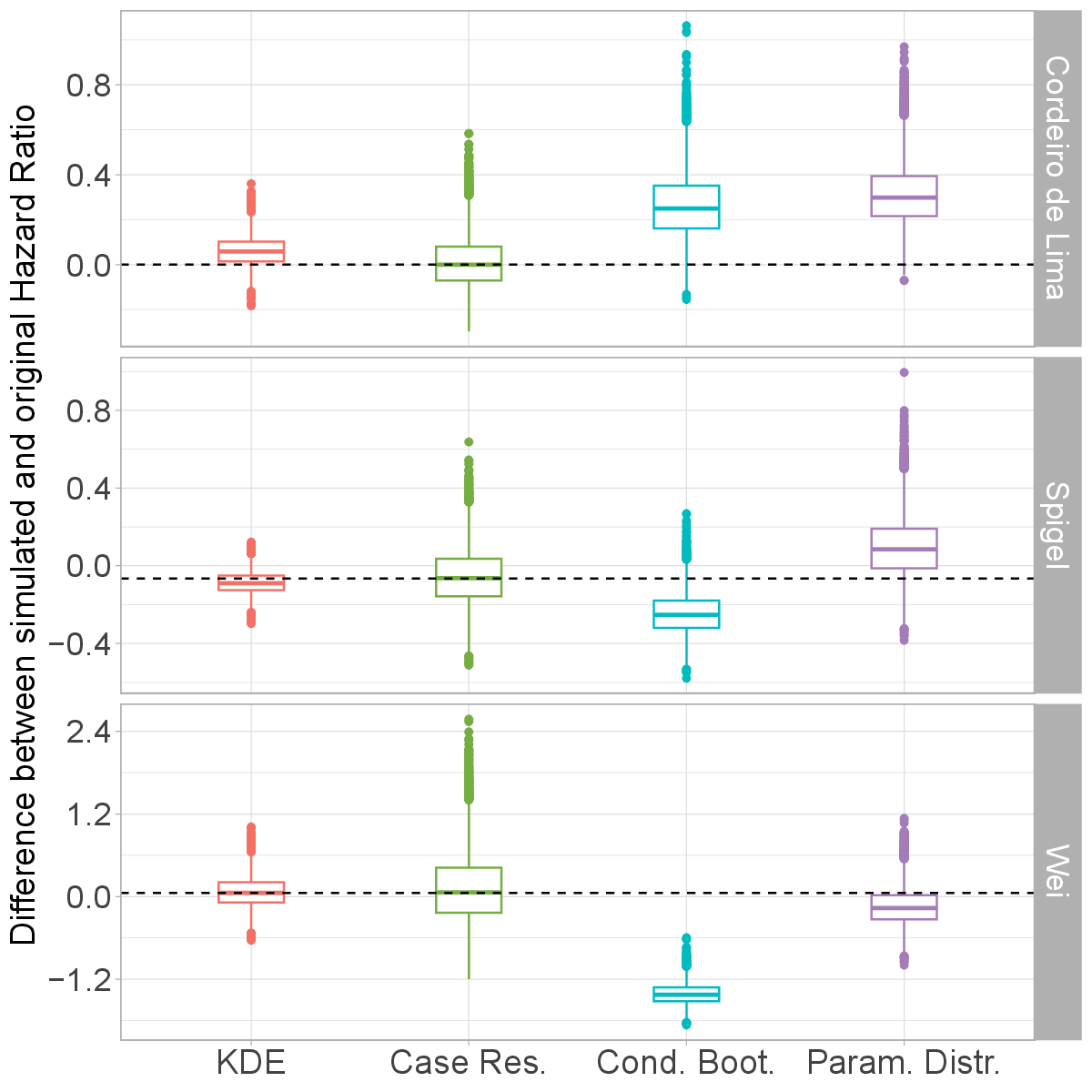}}
    \caption[Boxplots of the differences between the hazard ratios of the simulated data sets with crossing and non-crossing KM curves and the hazard ratios reported in the studies]{Boxplots of the differences between the hazard ratios of the simulated data sets with crossing (left) and non-crossing (right) KM curves and the hazard ratios reported in the studies. The dashed line is the difference between the hazard ratio of the reconstructed data set and the reported hazard ratio.}\label{fig:hr_cross_non_cross}
\end{figure}

For the studies with crossing KM curves, the boxplots of the described differences are displayed in Figure \ref{fig:hr_cross}.
The differences for the reconstructed data sets are very low. The maximum absolute difference for this group of studies is $0.010$ for the data set from Seto et al. This indicates that the hazard ratios were reconstructed well by the reconstruction algorithm.

For the data simulated by using the case resampling and the simulation based on kernel density estimation, the median hazard ratio is very similar to the reported hazard ratio for all of the studies. 
In the variability of the values, however, differences between the two simulation models can be observed. The values for the simulation based on kernel density estimation vary a bit less than the values for the case resampling.
When simulating the data based on the conditional bootstrap, the median hazard ratios for the simulated data sets for all of the studies are higher than the reported hazard ratios. For the study by Seto et al., the median difference is $0.010$ and slightly lower than for the kernel density estimation and the case resampling, but the variability is higher than for example for the kernel density estimation.
The median hazard ratios for the data sets simulated by using parametric distributions are similar to the reported hazard ratios and their variability is low.

The boxplots of the differences between the hazard ratios of the simulated and original data sets for the studies with non-crossing KM curves are shown in Figure \ref{fig:hr_non_cross}.
As for the studies with crossing KM curves, the hazard ratios of the reconstructed data sets are similar to the reported hazard ratios in the studies. 
In the simulation based on kernel density estimation, the median differences between the hazard ratios of the simulated and original data are similar to the differences between the hazard ratios of the reconstructed data sets and the reported hazard ratios and the observed differences show a very low variability.
The values for the case resampling, in contrast, show a higher variability than for the kernel density estimation while having a median value that is closer to the hazard ratio for the reconstructed data.
The differences between the hazard ratios for the data simulated by using the conditional bootstrap are in median $0.250$ higher than the reported value for the study by Cordeiro de Lima et al.
For the studies by Spigel et al. and Wei et al., the median hazard ratios for this simulation model are lower than the reported hazard ratio.
For the simulation based on parametric distributions, the results are a bit better than for the conditional bootstrap.
The variability of the observed values for the simulation based on parametric distributions is lower than, for example, the variability of the values for the conditional bootstrap but mostly higher than for the case resampling and the kernel density estimation.
The disparities between the results for the parametric distributions for the different studies might be explained by the distributions chosen for the respective studies. If the chosen distribution for a data set is not optimal, the results might be worse than when using the correct (unknown) distribution.

For the study with non-crossing KM curves and a late effect (Wu et al.), the results for the hazard ratios are similar to the results of Wei et al. Therefore, they are not further evaluated here. The boxplots of the differences between the hazard ratios of the simulated data sets and the reported values are displayed in Figure \ref{fig:hr_late} in the Appendix.

When considering the studies with crossing KM curves, the simulation models based on kernel density estimation, the case resampling and parametric distributions perform best with median differences close to zero and low variability. The hazard ratios for the simulation based on the conditional bootstrap deviate more from the reported values. For the studies with non-crossing KM curves, the simulation based on kernel density estimation and the case resampling perform best.
Therefore, for all study groups, the conditional bootstrap seems not sensible to be used. For non-crossing KM curves (with and without a visible late effect), the kernel density estimation or the case resampling should be used. For the studies with crossing KM curves, additionally, the simulation based on parametric distributions seems to be sensible.

In addition to the $p$-values of the logrank test and the hazard ratios, the restricted mean survival times and the median survival times were considered for the simulated data sets to assess how realistic the simulated data sets are. As the results do not differ much from the results for the $p$-values of the logrank test and the hazard ratios, they are not described in detail here. The description of the results for these measures can be found in the Appendix.
\subsection{Summary of the Results}
Regarding the reconstruction quality of the data, it can be said that, in general, the reconstructed data sets are similar to the original data.
Only the $p$-values of the logrank test show larger deviations for some methods, especially with the studies with crossing KM curves. A reason for this could be that the KM curves in these studies are sometimes very close to each other which makes digitizing the plots and thus reconstructing the data more difficult.

For the comparison of the simulation models, the runtimes of simulating a single data set, the $p$-values of the logrank test and the hazard ratios were considered.
When comparing the runtimes of the simulation models, it is noticeable that the simulation based on kernel density estimation has by far the highest runtimes (median runtime of $0.613$ seconds) among the four simulation models. The lowest runtimes can be observed for the case resampling (median runtime of $0.004$ seconds). The simulations based on the conditional bootstrap and the parametric distributions are just a bit slower than the case resampling.

The second criterion for the comparison of the simulation models is the $p$-values of the logrank test. For all of the studies and simulation models, there is a tendency for the $p$-values of the simulated data sets to be a bit higher than the $p$-values reported in the studies. The highest variability in the values and the highest deviations from the reported $p$-values can be observed for the simulation based on the conditional bootstrap. For the simulation based on the parametric distributions, the $p$-values are close to the reported $p$-values for the studies in which the KM curves are crossing.
The lowest variability and the $p$-values that are closest to the reported values can be observed for the simulation based on kernel density estimation. For the case resampling, the $p$-values are equally close to the reported values as for the kernel density estimation, but the variability is minimally higher.

For the hazard ratios, the results are similar to the results of the $p$-values. The highest accuracy was observed for the simulation based on the case resampling. While showing slight differences from the reported values, the second best-performing simulation model is the simulation based on kernel density estimation. For this method, the values vary less but the median hazard ratios show higher differences from the reported hazard ratios as for the case resampling. Just as with the $p$-values, the simulations based on the conditional bootstrap and the parametric distributions perform worse than the other two models.

Overall, regarding the results for the $p$-values and the hazard ratios, it can be concluded that the simulation based on kernel density estimation performs best. More accurate values with a higher variability can be observed for the case resampling. If the runtime of the different simulation models is taken into account as well, the case resampling performs better than the kernel density estimation.
For the simulation based on parametric distributions, the results vary a lot between the different studies. This indicates that most real world data do not follow one of the specific distributions considered here.

The results of the simulation study are summarized in Table \ref{tab:sum_results}.

\begin{table}[h]
	\centering
	\caption{Overview of the simulation results. A \checkmark indicates a good performance of the method, (\checkmark) refers to a good performance for only some settings/studies and a \ding{55} indicates a bad overall performance.\label{tab:sum_results}}
	\begin{tabular}{|l|c|c|c|}
		\hline &&&\\[-2.5ex]
		\textbf{Simulation Model} & \textbf{Runtime}&\textbf{$p$-value}&\textbf{Effect sizes}\\ \hline &&&\\[-2ex]
		Kernel Density Estimation &(\checkmark)&\checkmark&\checkmark\\[1ex]
		Case Resampling &\checkmark&\checkmark&\checkmark\\[1ex]
		Conditional Bootstrap &\checkmark& \ding{55}& \ding{55}\\[1ex]
		Parametric Distributions &\checkmark&(\checkmark)& \ding{55}\\[0.5ex] \hline
	\end{tabular}
\end{table}
\section{Discussion and Outlook}\label{sec:conclusion}
Benchmark data sets and simulation studies play an important role in model comparison in statistics. In the context of survival analysis and especially in oncology, there is a lack of adequate benchmark data sets. For simulation models, it is often not clear, how to simulate realistic data. Therefore, this paper has two goals: The first goal is to reconstruct data sets from published plots to provide them as benchmark data sets and the second goal is to compare the simulation based on parametric distributions, kernel density estimation and two resampling methods (case resampling and conditional bootstrap) in terms of their ability to simulate realistic data sets. The simulation is based on the previously reconstructed data sets. To assess how realistic the simulated data sets are, the $p$-values of the logrank test and the hazard ratios of the simulated data sets are compared to the reported values from the studies.

These studies were selected on the basis of previously defined criteria. For the reconstruction, it is important to have reported Kaplan-Meier (KM) curves and the number at risk at the different time points. Additionally, only two-sample comparisons in the field of lung cancer were selected. After adding some other criteria, like the resolution of the reported plots, ten studies are left, from which seven are selected for the simulation study. These can be divided into a group with crossing survival curves, a group with non-crossing survival curves and no late effect as well as a group with non-crossing curves and a visible late effect.

For the reconstruction, a widely used reconstruction algorithm is used. When comparing the 
median survival time, the $p$-values of the logrank test and the hazard ratios of the reconstructed and the original data, it can be said that the reconstruction is very accurate and the data sets can be used as benchmark data sets for the comparison of the simulation models. The highest deviations from the original values can be observed for the data sets from the group with crossing KM curves, but the reconstruction quality is still very good even for those data sets.

The simulation models were compared in a simulation study, in which $10{,}000$ data sets were simulated for each of the simulation models and studies. For the simulated data sets, the $p$-values of the logrank test and the hazard ratios are calculated. The differences between these values and the values reported in the studies are used to compare the different models. It is noticeable that for the $p$-values there is a tendency of minimally higher values in the simulated data sets. For the studies with crossing KM curves, the lowest differences from the reported values can be observed for the kernel density estimation and the case resampling. The highest deviations (with a high variability) can be noted for the conditional bootstrap. For the hazard ratios, the lowest differences between the reported values and the hazard ratios from the reconstructed data sets can be observed for the kernel density estimation and the case resampling. Just as for the $p$-values, the highest absolute differences can be observed for the conditional bootstrap. For the studies with non-crossing KM curves, the results are very similar. In addition to the kernel density estimation and the case resampling, the simulation based on parametric distributions performs very well. As an additional criterion, the runtime of the models can be considered. The lowest runtimes (for simulating a single data set) can be observed for the case resampling. The simulation based on kernel density estimation is in median more than $154.750$ times slower than the case resampling and $59.9$ times slower than the simulation based on parametric distributions. Since the simulation based on the case resampling has a high accuracy and a very low runtime, this can be considered a good method to simulate realistic survival data on the basis of an available (for example, benchmark) data set.

The results can be used as a suggestion for how (i) to create more benchmark data sets in the clinical field and (ii) to simulate more realistic survival data based on an available data set. Often, data sets are simulated by sampling from a specific distribution, which might not be a realistic representation of the actual distribution. The results of this study show that combining the two model comparison approaches by first reconstructing data sets from studies and then simulating data based on the reconstructed data is a good procedure for generating survival data.

Overall, based on the results of this paper, using benchmark data sets (for example reconstructed data sets) can be recommended as a basis for the simulation of realistic survival data. For the simulation, the case resampling or the slower kernel density estimation should preferably be used. This results in realistic data sets that can be used for model comparison and have the advantage that certain properties are known and a theoretically unlimited amount of data can be generated.

It should be noted that only three measures were considered descriptively for the comparison of the models in this study and that only studies on lung cancer patients were used. In further analyses, this could be extended to other measures, as for example the distance between the distributions of the simulated and reconstructed data sets, further simulation models, like plasmode simulations \cite{Franklin.2014}, could be compared to the simulation models considered in this study. Since the results might be different for other cancer types, other cancer types of cancer could be considered. In addition, besides only simulating the survival times, the simulation of realistic covariables for the patient data is of great importance. Due to the large number of potential settings with covariates and approaches to tackle these, these will be investigated in subsequent research. 
Moreover, the question as to why the $p$-values of the logrank test for the simulated data sets show the tendency to be a bit larger than the $p$-values reported in the studies could be further investigated. 

\section*{Acknowledgements}
We thank Anne-Laure Boulesteix (LMU) for many fruitful discussions on this work.
The authors gratefully acknowledge the computing time provided on the Linux HPC cluster at TU Dortmund University (LiDO3), partially funded in the course of the Large-Scale Equipment Initiative by the German Research Foundation (DFG) as project 271512359. The HPC cluster LiDO3 at TU Dortmund University is a heterogeneous computing cluster with 30\,TB RAM and 8,160 CPU cores spread over 366 nodes.

\noindent\textbf{Funding}\\
The work of Marc Ditzhaus, Markus Pauly and Maria Thurow was supported by the Deutsche Forschungsgemeinschaft (DFG, German Research Foundation) - PA 2409/5-2. Moreover, this work has been partly supported by the Research Center Trustworthy Data Science and Security (\url{https://rc-trust.ai}), one of the Research Alliance centers within the University Alliance Ruhr (\url{https://uaruhr.de}) and the DFG, grant number BO 3139/9-1.

\noindent\textbf{Conflicts of Interest}\\
The authors have declared no conflict of interest.

\bibliographystyle{vancouver}
\bibliography{literatur.bib}

\newpage

\renewcommand{\thesection}{A}
\numberwithin{figure}{subsection}
\numberwithin{table}{subsection}
\section{Appendix}
\renewcommand{\thesubsection}{A\arabic{subsection}}
\subsection{Information on the used Software Packages}\label{app:soft}
The reconstruction of the data sets, the simulation and the evaluation are performed using the statistical software \texttt{R} \cite{R2022} and the package \texttt{survival} \cite{survival-package}. For generating the plots, the packages \texttt{ggplot2} \cite{Wickham2016ggplot} and \texttt{survminer} \cite{Kassambara2021survminer} are used.

For some of the distributions used for simulating the data based on parametric distributions, the distribution functions are not implemented in the \texttt{stats} package in \texttt{R}. For the inverse gamma and the log-logistic distribution, the \texttt{R} package \texttt{actuar} \cite{Dutang2008} is used. The functions for the Gompertz distribution are taken from the \texttt{flexsurv} package \cite{Jackson2016flexsurv} and the functions for the gumbel distribution from the package \texttt{ordinal} \cite{Christensen2022ordinal} are used.

To estimate the parameters of the distributions by maximum likelihood estimation the following functions and \texttt{R} packages are used: the function \texttt{mlinvgamma} from the \texttt{R} package \texttt{univariateML} \cite{Moss2019univariateML} for the inverse gamma distribution and the function \texttt{fitdist} from the \texttt{R} package \texttt{fitdistrplus} \cite{Muller2015fitdistrplus} for all other distributions. In the \texttt{mlinvgamma} function the Newton-Raphson method is applied to find the maximum of the likelihood function while the \texttt{fitdist}-function uses the Nelder-Mead method for distributions with at least two parameters and the BFGS method for one-parametric distributions.

\subsection{Further Evaluations}\label{app:further_eval}
In the following chapter, further evaluations of the reconstructed and simulated data sets are conducted.
Besides the median survival time, which is also reported in the studies, the restricted mean survival time is considered here. An advantage of these two measures is that no proportional hazards assumption is needed.

\subsubsection*{Restricted Mean Survival Time}
The restricted mean survival time (RMST) is defined as the area under the KM curve of one group up to a certain time point $\tau \in [0,\infty)$ \cite{Royston2011RMST}. The \texttt{R} package \texttt{survRM2} \cite{Hajime2022survRM2} is used to calculate the RMST for the groups. In the \texttt{rmst2}-function of this package, the parameter $\tau$ is set to the maximum censoring time, if in exactly one of the treatment groups, the observation with the maximum time is censored or if the maxima of both groups are events. If in both groups, the maximum value belongs to a censored observation, $\tau$ is set to the lower of these two values.
In the following, the differences between the RMSTs of the two treatment groups are considered for the analysis. For simplicity, these differences are denoted by RMSTD in the following.

To compare the different simulation models, the differences between the RMSTDs of the simulated data sets and the reconstructed data sets are considered. For the data sets  simulated by using parametric distributions, the summary statistics of the RMSTD differences are displayed in Table \ref{tab:RMST_param}. The range of differences are very high for all of the considered studies. The inter quartile range is slightly lower. However, it can be seen in the table, that the 0 lies in none of the ranges between the 1st and 3rd quartile. This indicates that the RMSTD could not be reproduced well by this simulation model.

\begin{table}[h]
\caption{Summary statistics of the differences between the RMSTD of the simulated data sets and the RMSTD of the reconstructed data sets for the data sets simulated by using parametric distributions.}
    \centering
    \begin{tabular}{|l|c|c|c|c|c|c|}
    \hline
         \textbf{Study}&\textbf{Minimum} & \textbf{1st Quartile} & \textbf{Median} & \textbf{Mean} & \textbf{3rd Quartile} & \textbf{Maximum}\\\hline
         Cordeiro de Lima et al. \cite{CordeirodeLima.2018}&$-18295.98$& $-85.11$& $-38.69$& $-80.8$& $-13.38$&$229.49$\\\hline
         Liang et al. \cite{Liang.2017}&$-1298$&$11$&$22$&$423133$&$34$&$1472451526$\\\hline
         Seto et al. \cite{Seto.2020}&$-630.65$&$-7.26$&$8.69$&$112.07$&$42.23$&$159624.59$\\\hline
         Spigel et al. \cite{Spigel.2022}&$-24212.7$&$-76.42$&$-24.98$&$-77.26$&$14.62$&$21.01$\\\hline
         Wei et al. \cite{Wei.2020}&$-16501.592$&$10.278$&$21.013$&$-9.692$&$32.732$&$4106.899$\\\hline
         Wu et al. \cite{Wu.2015}&$-2923.173$&$-0.104$&$2.674$&$1.035$&$6.168$&$96.115$\\\hline
         Yoshioka et al. \cite{Yoshioka.2019}&$-1466.099$&$-29.297$&$-10.336$&$-27.015$&$-2.582$&$374.574$\\\hline
    \end{tabular}
    
    \label{tab:RMST_param}
\end{table}

\begin{figure}[!h]
    \centering
    \includegraphics[width = 0.9\linewidth]{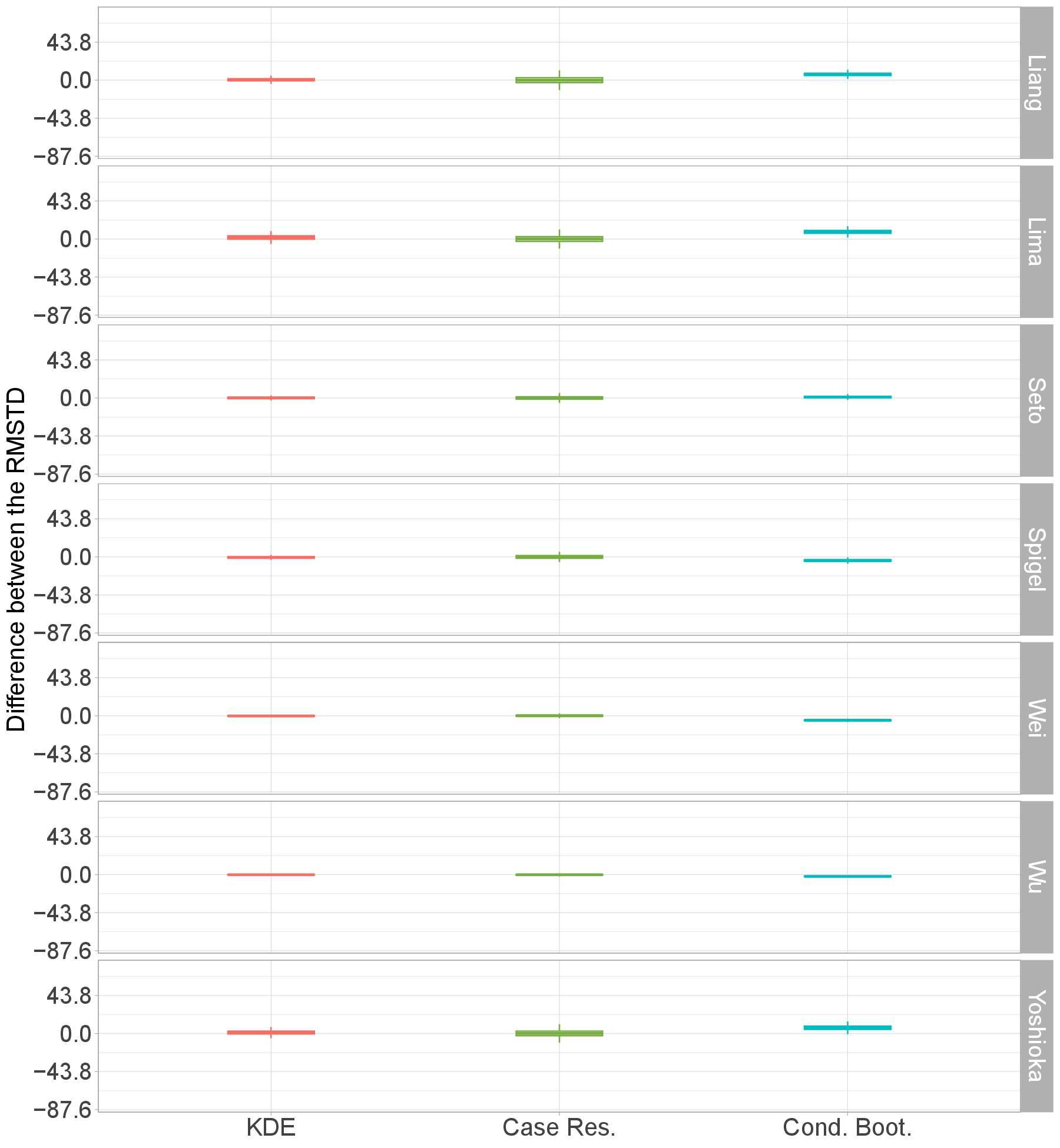}
    \caption{Boxplots of the differences between the RMSTD of the simulated data sets and the RMSTD of the reconstructed data sets for the data sets simulated by using parametric distributions.}
    \label{fig:RMST}
\end{figure}

In Figure \ref{fig:RMST}, the differences between the RMSTDs of the simulated data sets and the reconstructed data sets are displayed for the three remaining simulation models. Due to the extreme outliers, the simulation based on parametric distributions is not considered here. The values for the three remaining simulation models show a very low variability. For the simulation based on kernel density estimation and the case resampling, the median differences are very close to zero. For the conditional bootstrap, the median of the differences is close to zero for most studies as well. However, for some studies (e.g., the studies from Cordeiro de Lima et al. and Yoshioka et al.) the median differences are slightly larger respectively lower than zero. It still performs slightly worse than the kernel density estimation and the case resampling but much better than the simulation based on parametric distributions.

\pagebreak
\newpage
\subsubsection*{Median Survival Time}

For the data set from Wei et al., no median survival time could be computed for the original data in the study for the MWA/Chemotherapy group. For the simulated data sets, the median survival time could be calculated for $4{,}169$ of $10{,}000$ iterations for the simulation based on kernel density estimation, for $4{,}428$ data sets simulated by using the case resampling, in all $10{,}000$ iterations of the conditional bootstrap and for $9{,}994$ of $10{,}000$ data sets for the simulation based on parametric distributions. A reason for this could be, that the estimated survival probability for this treatment group is slightly higher than $0.5$ at the end of the study.

In addition, for the study from Wu et al., for some simulation models and iterations, no median survival time can be calculated. For the simulation based on kernel density estimation, for $89$ of the $10{,}000$ iterations, no median survival time can be computed. For the case resampling, this is the case for $293$ data sets and for the simulation based on parametric distributions, the median survival time cannot be calculated for $14$ of $10{,}000$ simulated data sets. For the conditional bootstrap, the median survival times can be calculated for all $10{,}000$ simulated data sets.

\begin{figure}[!h]
	\centering
	\includegraphics[width = 0.95\linewidth]{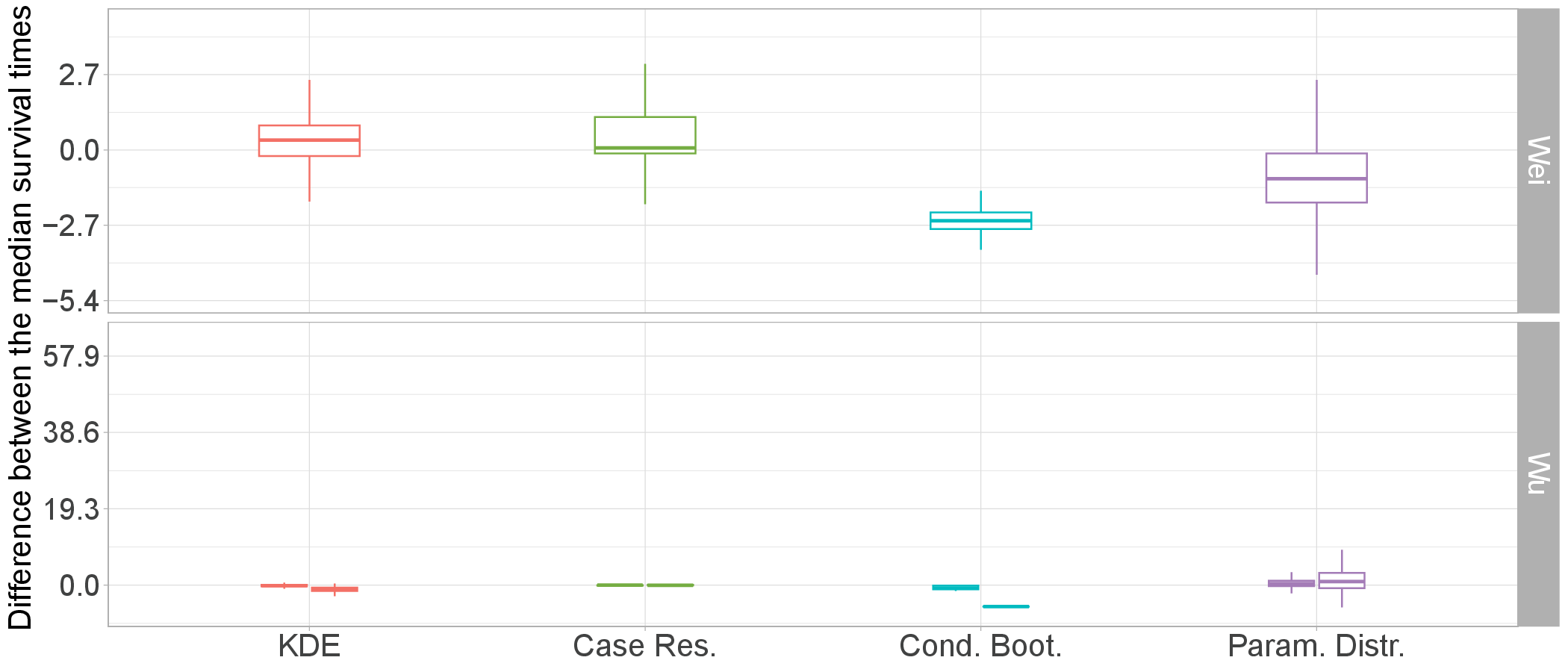}
	\caption{Boxplots of the differences between the median survival times of the treatment groups of the simulated data sets and the median survival time of the treatment groups of the reconstructed data sets for the studies from Wei et al. and Wu et al.\label{fig:med_ST_wrongNA}}
\end{figure}

In Figure \ref{fig:med_ST_wrongNA}, the boxplots of the differences between the median survival times of the simulated data sets and the median survival times in the studies are displayed. The pairs of boxplots refer to the two treatment groups. Since for the study from Wei et al., the survival time is not reported for one of the treatment groups, only one boxplot is shown for each of the simulation models. For the study of Wu et al., no median survival time can be calculated in some iterations. This should be noted when comparing the simulation models. For both studies, the median difference is very close to zero for the case resampling and only slightly higher for the kernel density estimation. The conditional bootstrap shows high deviations from the reported median survival times and the simulation based on parametric distributions performs only slightly better.

\begin{figure}[!h]
	   \centering
	   \includegraphics[width = 0.95\linewidth]{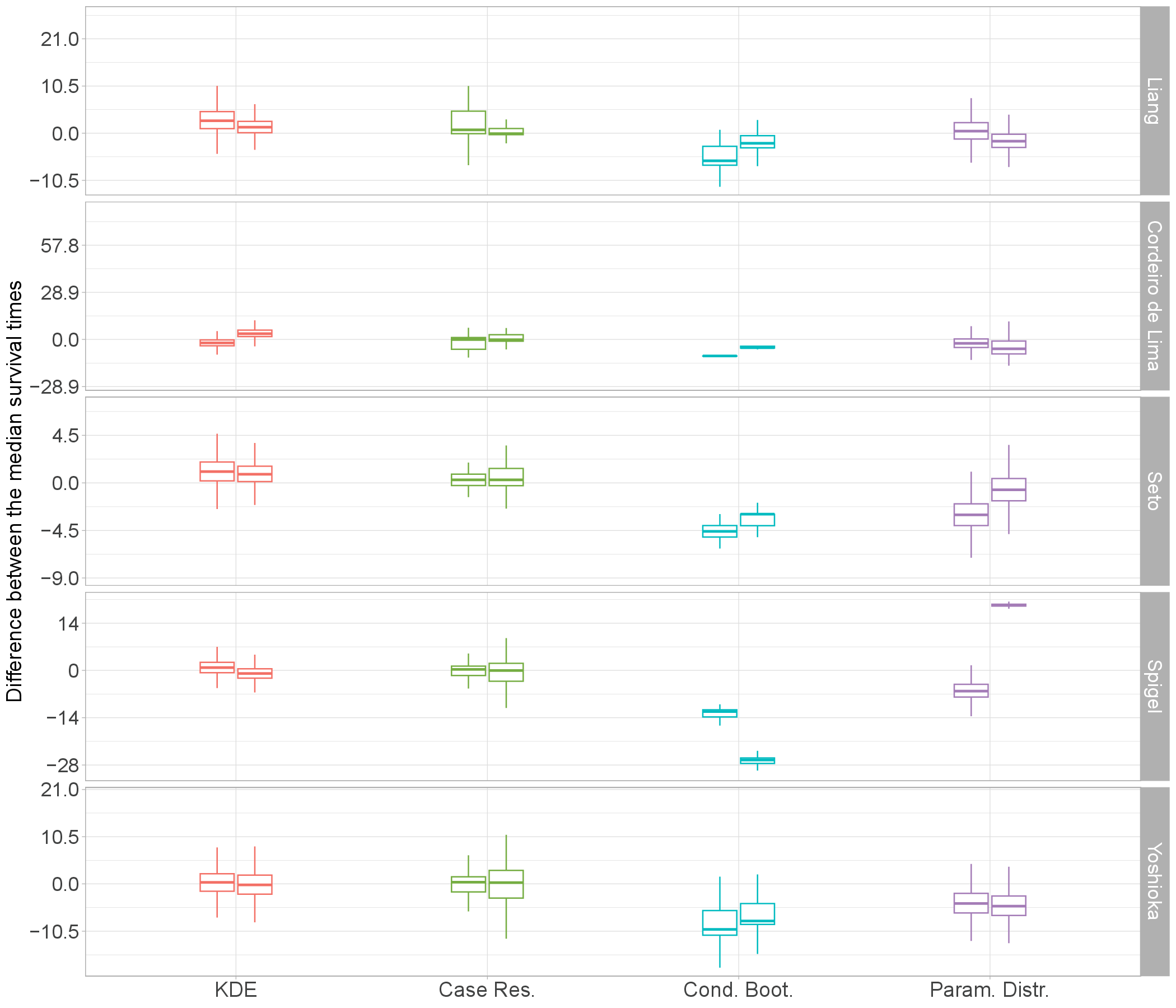}
	   \caption{Boxplots of the differences between the median survival times of the treatment groups of the simulated data sets and the median survival time of the treatment groups of the reconstructed data sets for the studies from Liang et al., Cordeiro de Lima et al., Seto et al., Spigel et al. and Yoshioka et al.\label{fig:med_ST_nowrongNA}}
    \end{figure}

When considering the studies, for which no faults were made regarding the calculability of the median survival times, the results are similar. The boxplots of the differences between the median survival times of the simulated data sets and the median survival times in the studies are shown in Figure \ref{fig:med_ST_nowrongNA}. For the case resampling and the simulation based on kernel density estimation, the median survival times are in median very close to the reported median survival times and for the conditional bootstrap and the simulation based on parametric distributions, high deviations can be observed for some studies.

\clearpage
\subsection{Tables} \label{app:tab}
\begin{table}[h]
	\centering 
	\caption[Comparison of the median survival time in the groups of the original and reconstructed data sets and the $p$-value of the logrank test]{Comparison of the median survival time in the groups of the original and reconstructed data sets and the $p$-value of the logrank test. \label{tab:recon_comp}}
	\begin{tabular}{|l|c|c|c|c|c|}
		\hline
		\multirow{2}{*}{\textbf{Study}} &  \multicolumn{2}{c|}{\textbf{Logrank test $p$-value}} &\multirow{2}{*}{\textbf{Treatment Group}}&\multicolumn{2}{c|}{\textbf{Median Survival Time}}\\ \cline{2-3}\cline{5-6}
		&\textbf{Reconstructed}&\textbf{Reported}&&\textbf{Reconstructed}&\textbf{Reported}\\\hline
		\multirow{2}{*}{Liang et al. \cite{Liang.2017}} & \multirow{2}{*}{$0.103$ }&\multirow{2}{*}{$0.095$}& EP& 24 &23.3\\ \cline{4-6}
		&&& PC& 20.6 &20.7\\ \hline
		\multirow{2}{*}{Cordeiro de Lima et al. \cite{CordeirodeLima.2018}} & \multirow{2}{*}{$0.008$ }&\multirow{2}{*}{$0.009$}& $<61 \text{Gy}$& 15.9 &16\\ \cline{4-6}
		&&& $\geq 61 \text{Gy}$& 25.9 &26\\  \hline
		\multirow{2}{*}{Seto et al. \cite{Seto.2020}} &\multirow{2}{*}{$0.171$}&\multirow{2}{*}{$0.069$}& Pem + Bev&23.6 &23.3\\ \cline{4-6}
		&& &Bev&19.9 &19.6\\  \hline
		\multirow{2}{*}{Spigel et al. \cite{Spigel.2022}} & \multirow{2}{*}{$0.001$}&\multirow{2}{*}{$0.003$}& Placebo& 29.4 &29.1\\ \cline{4-6}
		& & &Durvalumab& 47.5 &47.5\\  \hline
		\multirow{2}{*}{Wei et al. \cite{Wei.2020}} & \multirow{2}{*}{$<0.001$}& \multirow{2}{*}{$< 0.001$}& MWA/Chemotherapy & \texttt{NA} &\texttt{NA}\\ \cline{4-6}
		& &&Chemotherapy&12.5 &12.4\\  \hline
		\multirow{2}{*}{Wu et al. \cite{Wu.2015}} &\multirow{2}{*}{$<0.001$} &\multirow{2}{*}{$<0.001$ } & GP&5.6 &5.6\\ \cline{4-6}
		& & &Erlotinib&11 &11\\  \hline
		\multirow{2}{*}{Yoshioka et al. \cite{Yoshioka.2019}} & \multirow{2}{*}{$0.231$}&  \multirow{2}{*}{$0.207$}& CD& 37.7 &37.3\\ \cline{4-6}
		&&& Gefitinib&35.2&34.9\\  \hline
	\end{tabular}
\end{table}

\clearpage
\subsection{Figures}\label{app:fig}

\begin{figure}[h]
\centering
\subfloat[][Seto et al. \cite{Seto.2020}]{\label{fig:recon_seto}\includegraphics[width = 0.48\linewidth, valign = b]{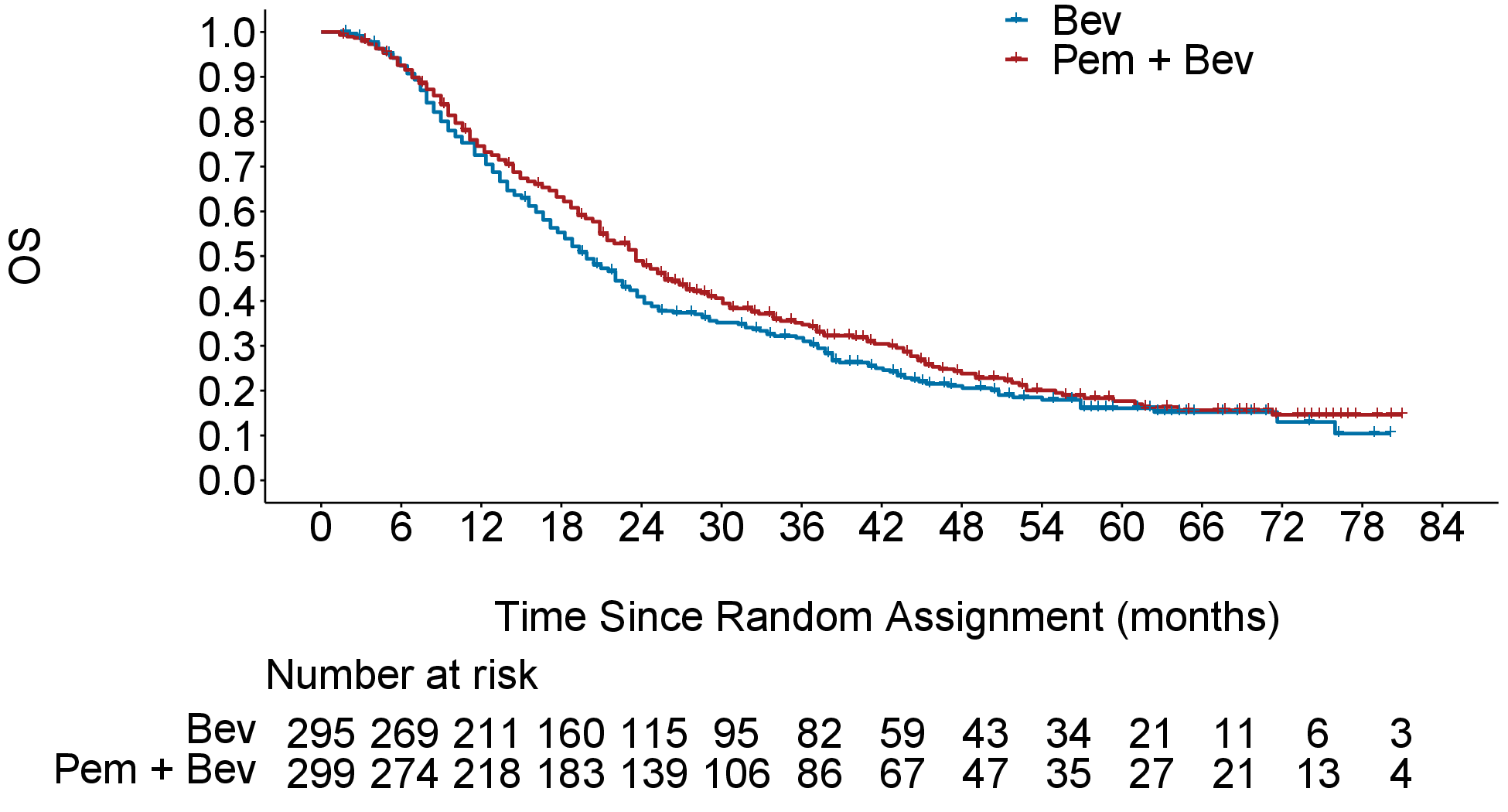}}
\hfill
\subfloat[][Spigel et al. \cite{Spigel.2022}]{\label{fig:recon_spigel}\includegraphics[width = 0.48\linewidth, valign = b]{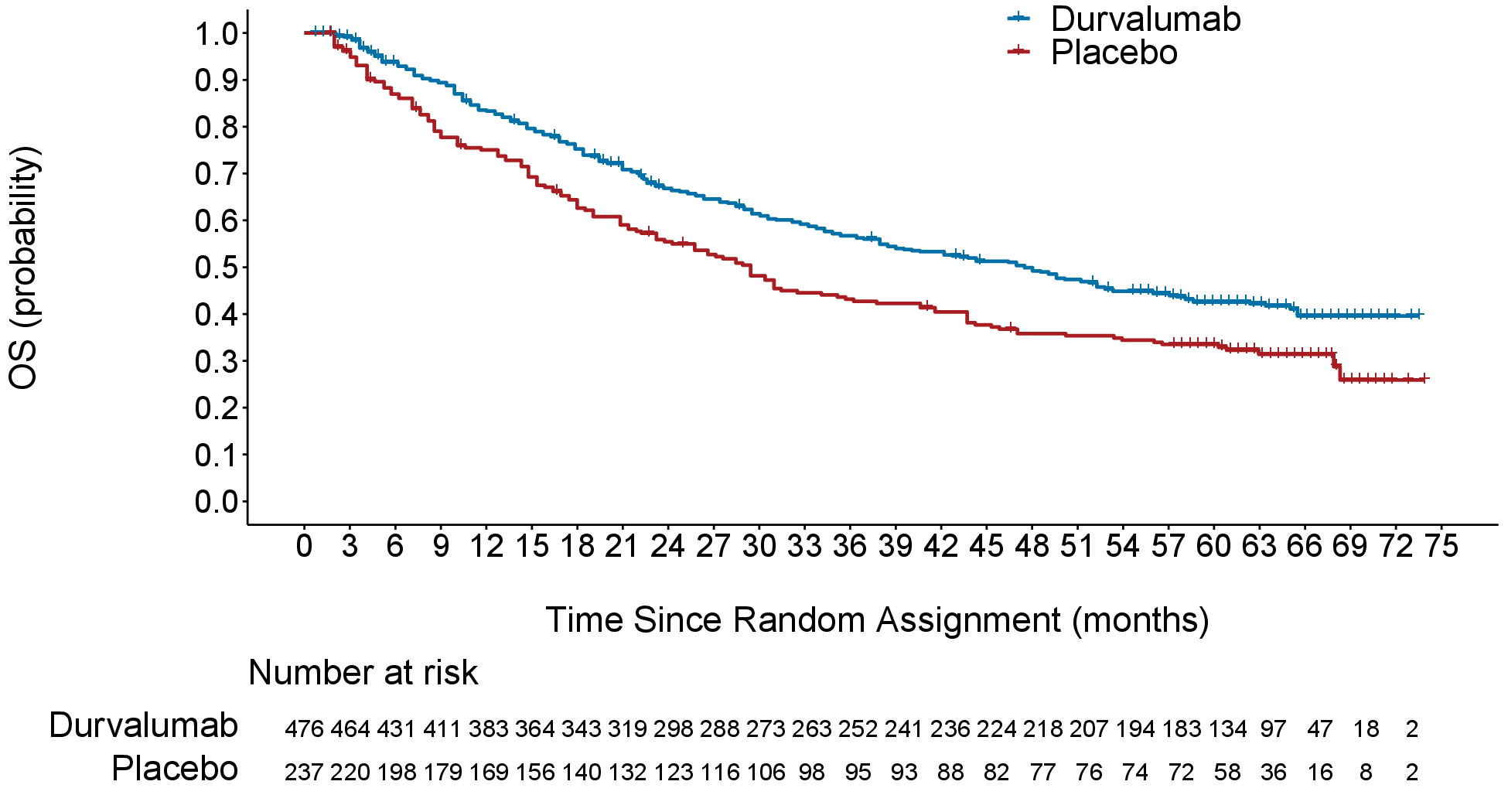}}\\[5ex]
\subfloat[][Wei et al. \cite{Wei.2020}]{\label{fig:recon_wei}\includegraphics[width = 0.48\linewidth, valign = b]{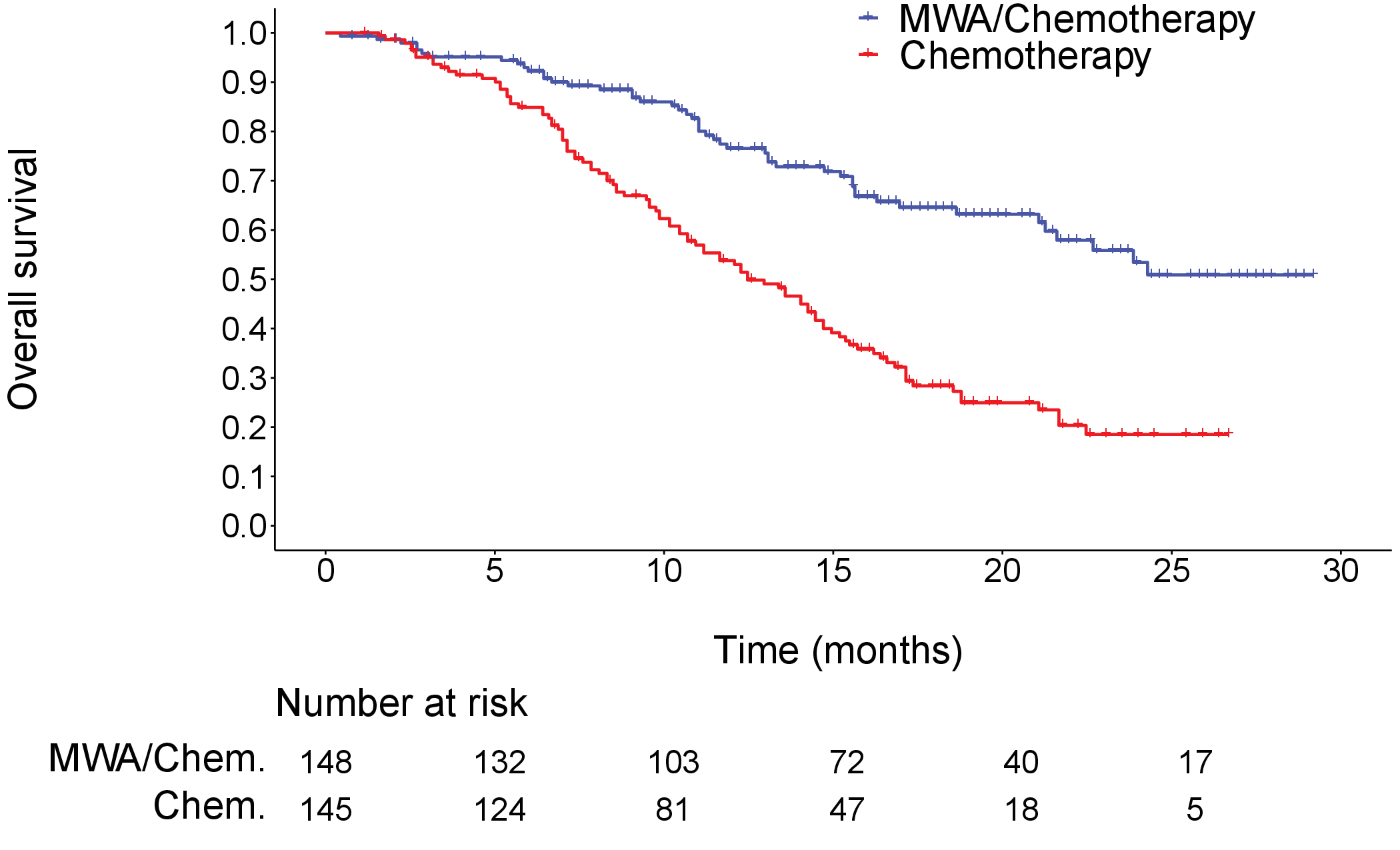}}
\hfill
\subfloat[][Yoshioka et al. \cite{Yoshioka.2019}]{\label{fig:recon_yoshioka}\includegraphics[width = 0.48\linewidth, valign = b]{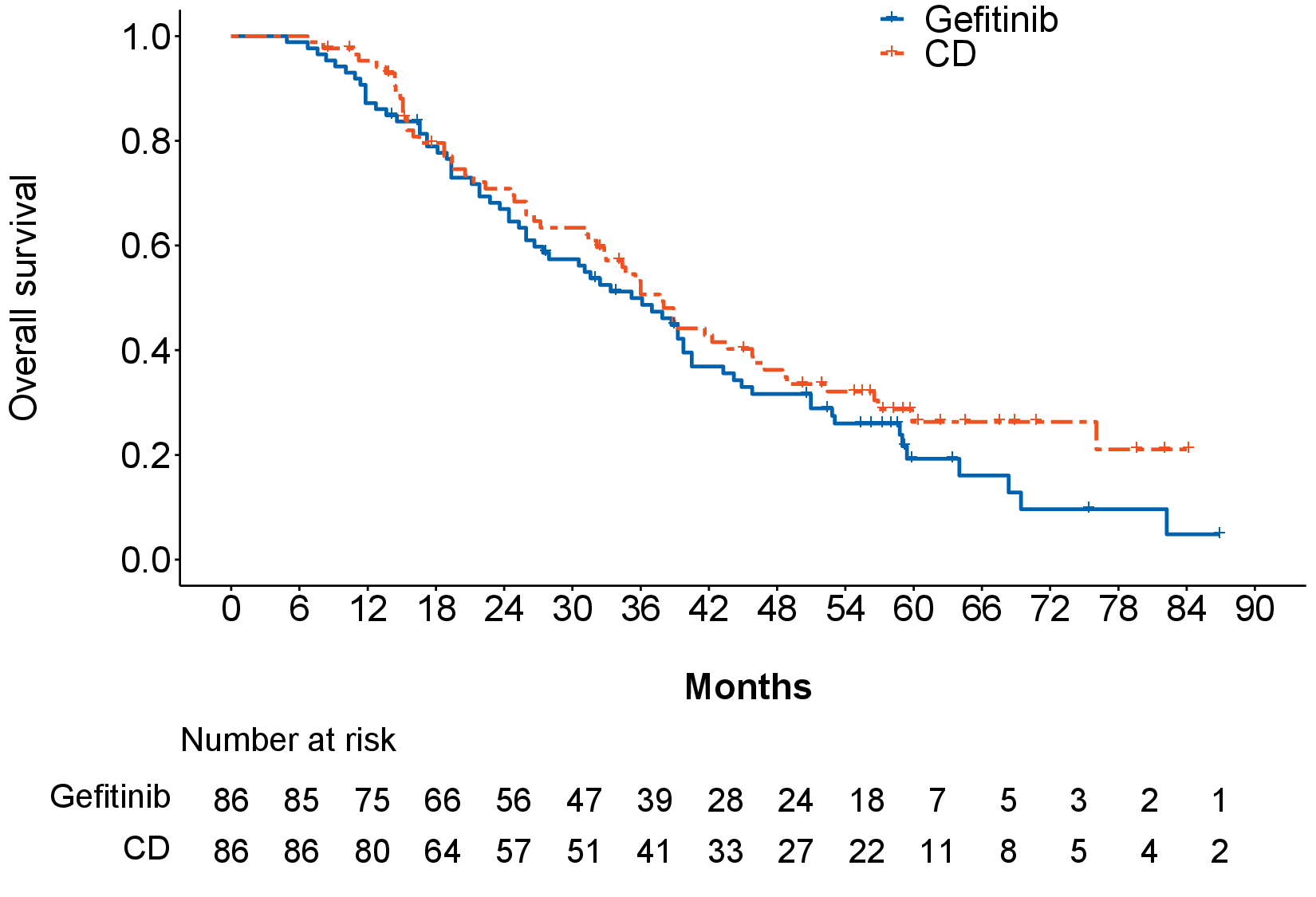}}
\caption{Kaplan-Meier curves of the reconstructed data sets.}
\label{fig:Reconstruction_appendix}
\end{figure}

\begin{figure}[!htbp]
	\centering
	\includegraphics[width = 0.9\linewidth]{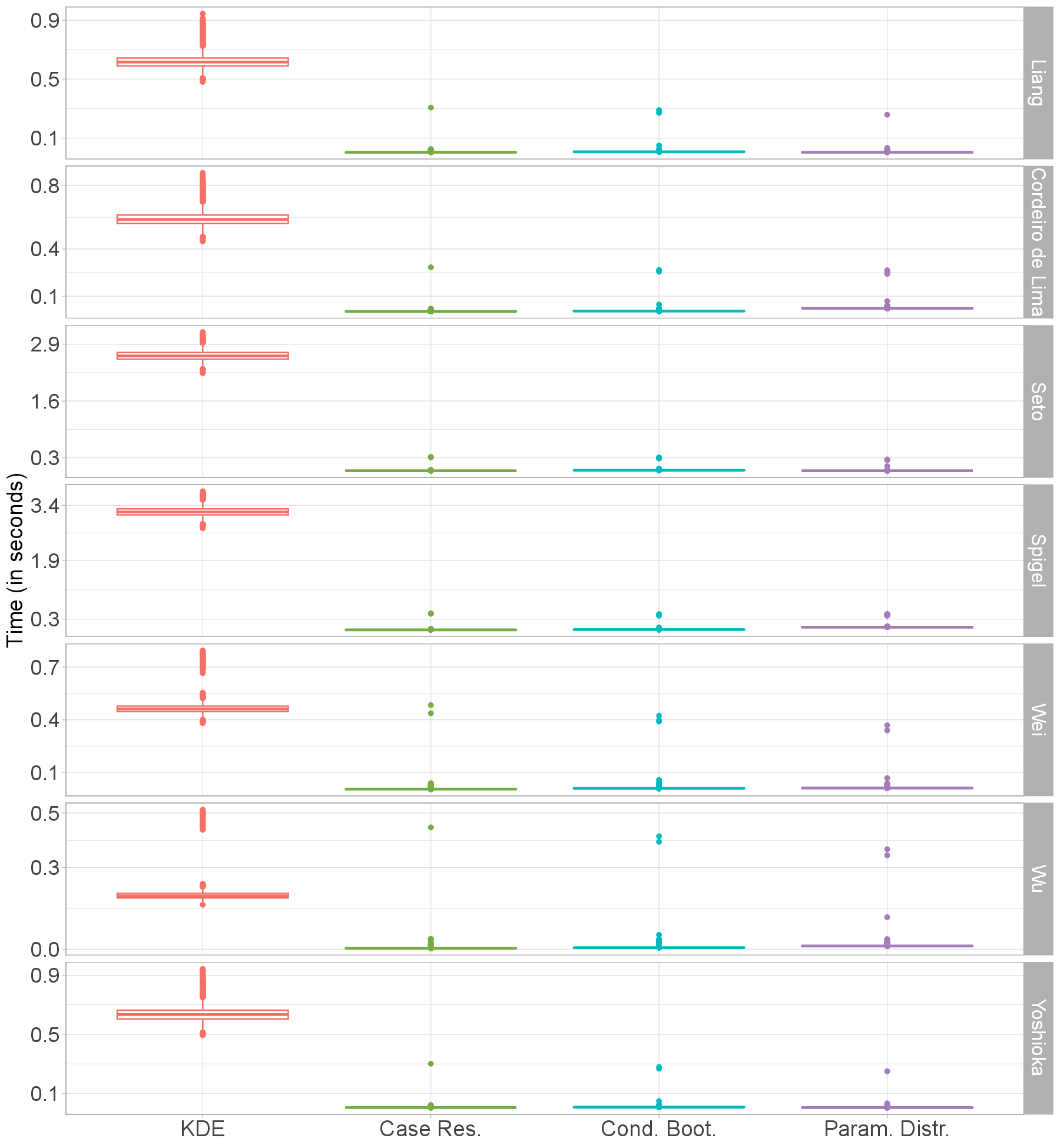}
	\caption{Boxplots of the runtimes of the simulation models for the seven studies.\label{fig:runtimes}}
\end{figure}

\begin{figure}[!h]
	\centering
	\includegraphics[width = 0.9\linewidth]{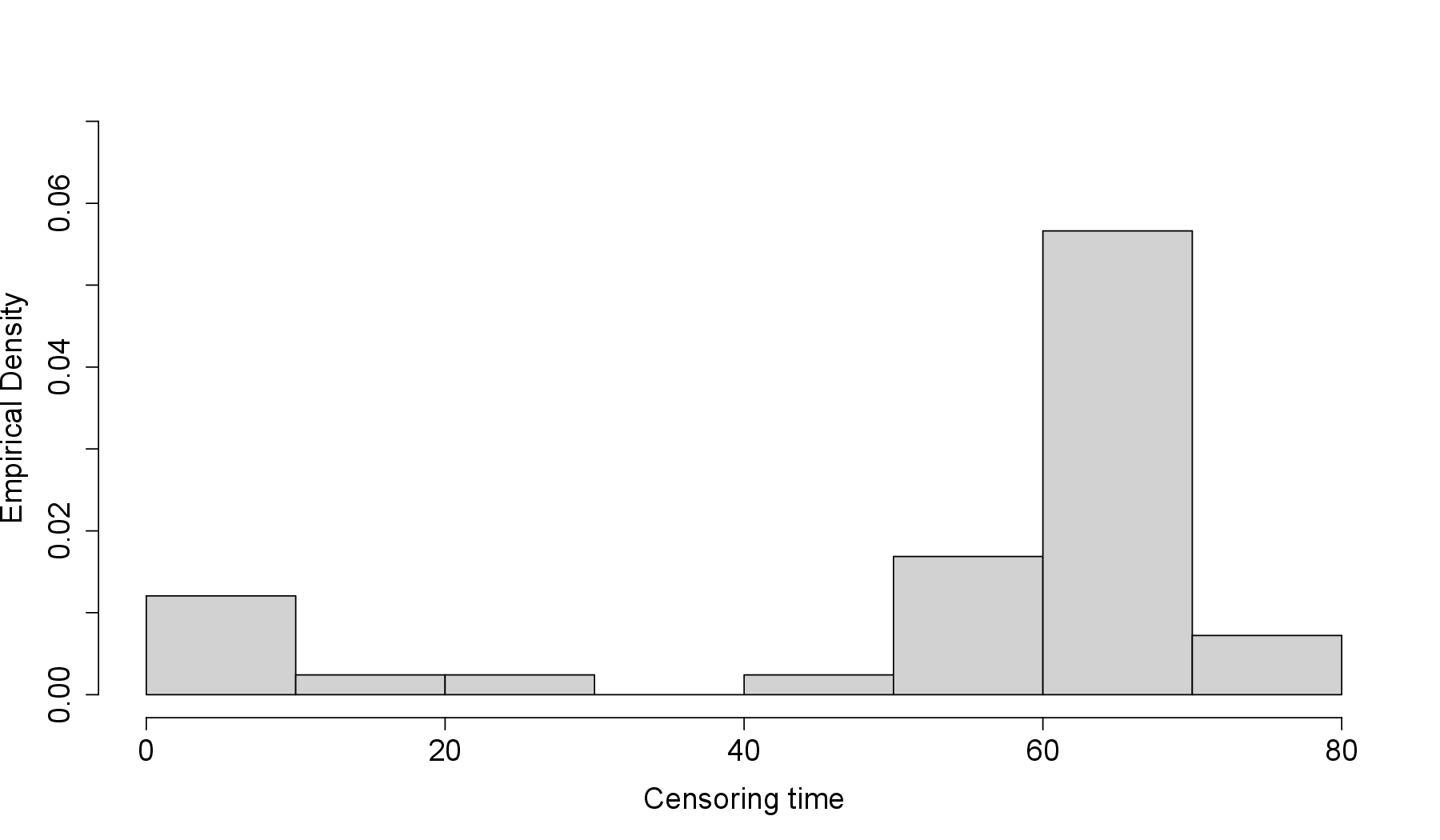}
	\caption{Histogram of the reconstructed censoring times in the placebo group of the study from Spigel et al.}\label{fig:hist_mixture}
\vspace*{4ex}
	\centering
	\includegraphics[width = 0.88\linewidth]{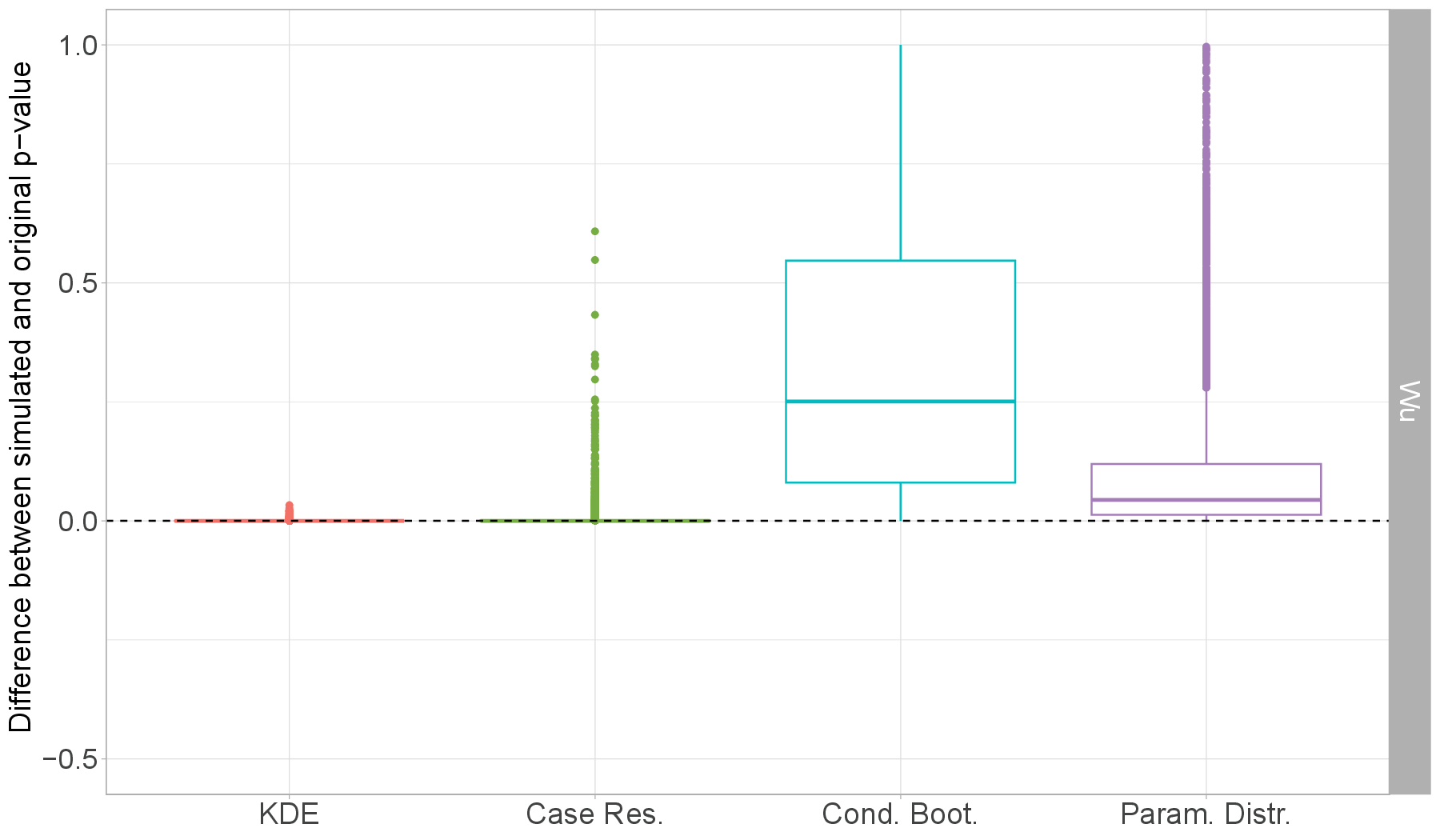}
	\caption[Boxplots of the differences between the $p$-values of the logrank test of the simulated data sets with non-crossing Kaplan-Meier curves with a late effect and the $p$-values reported in the studies]{Boxplots of the differences between the $p$-values of the logrank test of the simulated data sets with non-crossing KM curves with a late effect and the $p$-values reported in the studies. The dashed line is the difference between the $p$-value of the reconstructed data set and the reported $p$-value.}\label{fig:lr_p_late}
	\vspace*{\floatsep}
\end{figure}

\begin{figure}[!h]

		\centering
		\includegraphics[width = 0.88\linewidth]{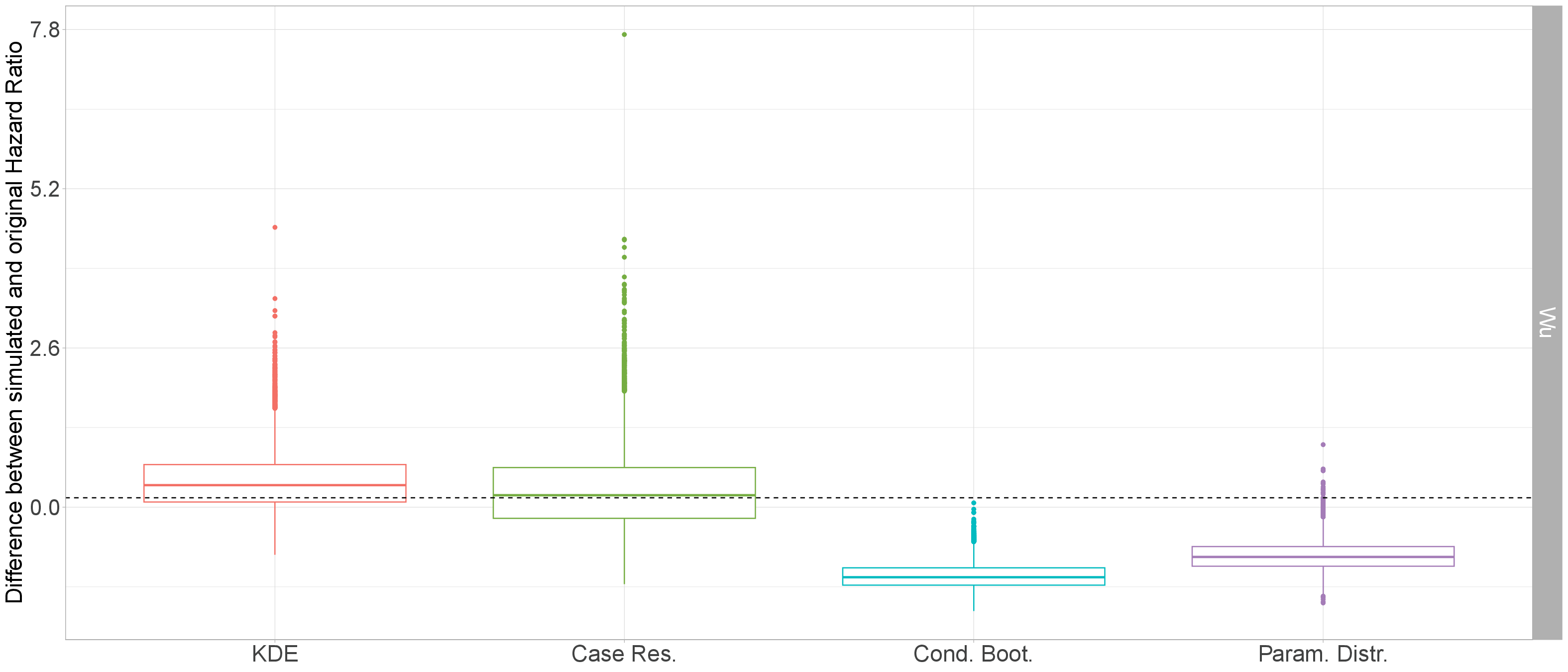}
		\caption[Boxplots of the differences between the hazard ratios of the simulated data sets with non-crossing KM curves with a late effect and the hazard ratios reported in the studies]{Boxplots of the differences between the hazard ratios of the simulated data sets with non-crossing KM curves with a late effect and the hazard ratios reported in the studies. The dashed line is the difference between the hazard ratio of the reconstructed data set and the reported hazard ratio.}\label{fig:hr_late}
\end{figure}

\begin{figure}[!h]
		\centering
		\includegraphics[width = 0.9\linewidth]{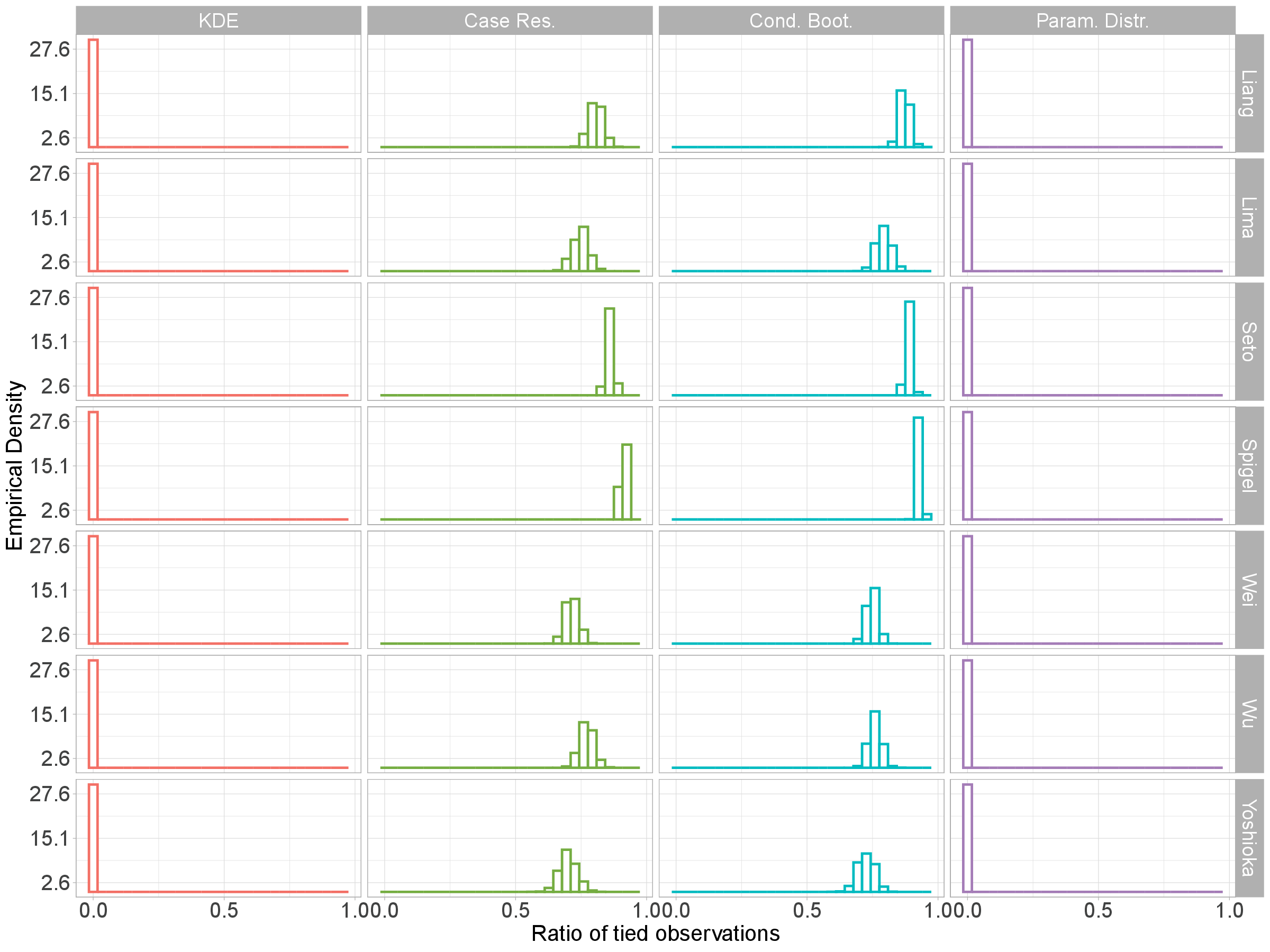}
		\caption[]{Histograms of the ratio of tied observations in the simulated data sets.}\label{fig:tie_ratio}
\end{figure}

\begin{figure}[!h]
		\centering
		\includegraphics[width = 0.9\linewidth]{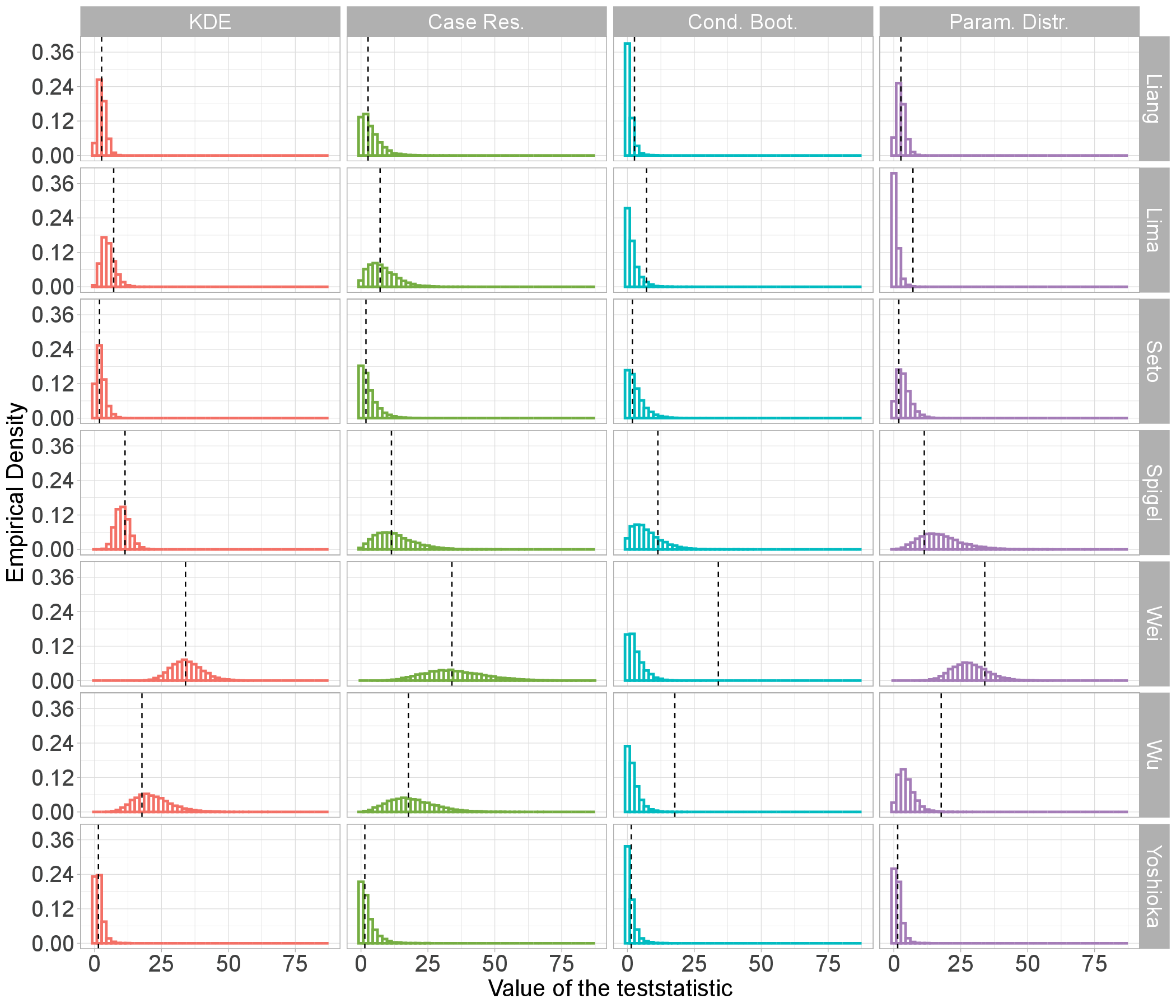}
		\caption[]{Histograms of the test statistics of the logrank test for the simulated data sets. The test statistics for the reconstructed data sets are displayed as dashed vertical lines.}\label{fig:teststats}
\end{figure}

\end{document}